\documentclass[11pt]{article}
\usepackage{amsfonts}
\usepackage{amsthm,amssymb,amsmath}

\newtheorem{theorem}{Theorem}

\numberwithin{equation}{section} \numberwithin{theorem}{section}
\newtheorem{proposition}[theorem]{Proposition}
\newtheorem{lemma}[theorem]{Lemma}
\newenvironment{remark}{\begin{aremark}\rm}{\end{aremark}}
\newtheorem{aremark}[theorem]{Remark}

\begin{document}

\title{Bulk Universality and Related Properties of Hermitian Matrix Models}
\author{L.Pastur and M.Shcherbina \\
Institute for Low Temperatures, Kharkiv, Ukraine}
\date{}
\maketitle

\begin{abstract}
\noindent We give a new proof of universality properties in the bulk of
spectrum of the hermitian matrix models, assuming that the potential that
determines the model is globally $C^{2}$ and locally $C^{3}$ function (see
Theorem \ref{t:U.t1}). The proof as our previous proof in \cite{Pa-Sh:97} is
based on the orthogonal polynomial techniques but does not use asymptotics
of orthogonal polynomials. Rather, we obtain the $sin$-kernel as a unique
solution of a certain non-linear integro-differential equation that follows
from the determinant formulas for the correlation functions of the model. We
also give a simplified and strengthened version of paper \cite{BPS:95} on
the existence and properties of the limiting Normalized Counting Measure of
eigenvalues. We use these results in the proof of universality and we
believe that they are of independent interest.
\end{abstract}


\section{Introduction}

We present an asymptotic analysis of a class of random matrix ensembles,
known as matrix models. They are defined by the probability law

\begin{equation}
P_{n,\beta }(d_{\beta }M)=Z_{n,\beta }^{-1}\exp \left\{ -\frac{\beta n }{2}%
\hbox{Tr}\,\,V(M)\right\} d_{\beta }M,  \label{MMb}
\end{equation}%
where $M=\{M_{jk}\}_{j,k=1}^{n}$ is a $n\times n$ real symmetric ($\beta =1$%
) or hermitian ($\beta =2$) matrix, $V:\mathbb{R}\rightarrow \mathbb{R}_{+}$
is a continuous function called the potential of the model and we will
assume that
\begin{equation}
V(\lambda )\geq 2(1+\epsilon )\log {(1+|\lambda |)}  \label{cond0}
\end{equation}%
for some $\epsilon >0$,
\begin{equation}
d_{1}M=\prod_{1\leq j\leq k\leq n}dM_{jk},\quad
d_{2}M=\prod_{j=1}^{n}dM_{jj}\prod_{j<k}d\Im M_{jk}d\Re M_{jk},  \label{dMb}
\end{equation}%
and $Z_{n,\beta }$ is the normalization constant.

These ensembles have been actively studied in the last decades because of
the number of their interesting properties and applications (see review
works \cite{DiF-Co:95,We-Co:98,Ka-Sa:99,Me:91} and references therein).

The Random Matrix Theory deals with several asymptotic regimes of
the eigenvalue distribution, in particular, the global regime,
centered around
the weak convergence of the Normalized Counting Measure of eigenvalues (see %
\ref{NCM}), and the local regime, where universality of local eigenvalue
statistics is one of the main topics. Universality of various ensembles of
hermitian and other matrices have important applications (see \cite%
{We-Co:98,Ka-Sa:99,Me:91}) and have been discussed in physics literature
since the beginning of modern era of Random Matrix Theory in the early
fifties \cite{Br-Ze:93,Dy:72,Ha-We:95,Ka-Co:88,Me:91,Mi-Fy:91,Mo:90,Wi:57}.
Rigorous proofs of the universality property for the hermitian matrix models
($\beta=2$) were given in \cite{Pa-Sh:97} and \cite{De-Co:99}. Both proofs
rely strongly on the orthogonal polynomial techniques, reducing the proof to
a certain asymptotic problem (see relation (\ref{U.1.23}) below) for a
special class of orthogonal polynomials. The reduction is based on
remarkable formulas for all marginals of the joint probability density of
eigenvalues known as determinant formulas (see formula \ref{pnlb} below) for
$\beta =2$.

In this paper we give a new proof of the bulk universality of local regime
of hermitian matrix models. The proof is valid for potentials in (\ref{MMb})
that are of the class $C^2$ everywhere and have 3 bounded derivatives in a
neighborhood of a point, where we prove the universality. We obtain the $%
\emph{sin}$-kernel as a unique solution of a certain nonlinear
integro-differential equation, while in our previous paper \cite{Pa-Sh:97}
the kernel was obtained, roughly speaking, as a power series in its
arguments. Since our proof of universality requires a number of facts on
limiting Normalized Counting Measure of eigenvalues of matrix models, the
paper includes an updated and simplified version of results of \cite{BPS:95}
on the existence and properties of the measure. Most of simplifications are
possible because of systematic use of book \cite{Sa-To:97}

The paper is organized as follows. In Section 2 we treat the global regime
and in Section 3 the local regime. In the course of our presentation we will
need several technical results from \cite{BPS:95,Pa-Sh:97}. We will give
them here (often improving) to make the paper self consistent.

\section{Global regime}

\subsection{Generalities}

Denote $\{\lambda _{l}^{(n)}\}_{l=1}^{n}$ the eigenvalues of a real
symmetric or hermitian matrix $M$ and set for any interval $\Delta \in
\mathbb{R}$
\begin{equation}
N_{n}(\Delta )=\sharp \{\lambda _{l}^{(n)}\in \Delta ,l=1,\dots ,n\}/n.
\label{NCM}
\end{equation}%
This is the Normalized Counting Measure of eigenvalues of $M$ (empirical
distribution in mathematical statistics). In this section we study the
convergence of the random measure $N_{n}$ to a non random limit $N$ which
proved to be a probability measure ($N(\mathbb{R})=1$) called often the
Integrated Density of States.

Our starting point is the joint probability density of eigenvalues,
corresponding to (\ref{MMb}) -- (\ref{dMb}) \cite{Me:91}.
\begin{equation}
p_{n,\beta }(\lambda _{1},...,\lambda _{n})=Q_{n,\beta }^{-1}\exp \left\{ -%
\frac{\beta n }{2}\sum_{i=1}^{n}V(\lambda _{i})\right\} |\Delta _{n}(\Lambda
)|^{\beta },  \label{psymb}
\end{equation}%
where
\begin{equation}
\Delta _{n}(\Lambda )=\prod_{1\leq j<k\leq n}(\lambda _{i}-\lambda
_{j}),\quad \Lambda =(\lambda _{1},...,\lambda _{n}),  \label{Delta}
\end{equation}%
and $Q_{n,\beta }^{-1}$ is the normalization constant.

Let
\begin{equation}
p_{l,\beta }^{(n)}(\lambda _{1},...,\lambda _{l})=\int_{\mathbb{R}%
^{n-l}}p_{n,\beta }(\lambda _{1},...\lambda _{l},\lambda _{l+1},...,\lambda
_{n})d\lambda _{l+1}...d\lambda _{n}  \label{pnlb}
\end{equation}%
be the $l$-th marginal density of $p_{n,\beta }$. Then, in particular,
\begin{equation}
\overline{N}_{n}(\Delta ):=E\{N_{n}(\Delta )\}=\int_{\Delta }p_{1,\beta
}^{(n)}(\lambda _{1})d\lambda _{1},  \label{Nbp}
\end{equation}%
or
\begin{equation}
\overline{N}_{n}(\Delta )=\int_{\Delta }\rho _{n}(\lambda )d\lambda ,\quad
\rho _{n}=p_{1,\beta }^{(n)}.  \label{Nnrho}
\end{equation}%
The cases $\beta =1$ and $\beta =2$ correspond to real symmetric and
hermitian matrices. However, the probability density (\ref{psymb}) is well
defined for any $\beta >0$ (in particular, the case $\beta =4$ corresponds
to real quaternion matrices \cite{Me:91}). In this section we will treat the
general case of $n$-independent strictly positive $\beta $.

According to Wigner (see \cite{Wi:57,Dy:62,Me:91}) the density (\ref{psymb})
can be written as the density of the canonical Gibbs measure
\begin{equation}
p_{n,\beta }(\Lambda )=Q_{n,\beta }^{-1}e^{-\beta n H(\Lambda )/2} ,\quad
\Lambda =(\lambda _{1},\dots ,\lambda _{n})\in \mathbb{R}^{n},
\label{IDS.1.6}
\end{equation}%
corresponding to a one-dimensional system of $n$ particles with the
Hamiltonian
\begin{equation}
H(\Lambda )=\sum_{i=1}^{n}V(\lambda _{i})-{\frac{1}{n}}\sum_{i<j}\log
|\lambda _{i}-\lambda _{j}|,  \label{IDS.1.7}
\end{equation}%
the temperature $2/\beta n$, and the partition function
\begin{equation}
Q_{n,\beta }=\int_{\mathbb{R}^{n}}e^{-\beta n H(\Lambda )/2}d\Lambda .
\label{Qnb}
\end{equation}%
The first term of the r.h.s. of (\ref{IDS.1.7}) is analogous to the energy
of particles due to the external field $V$ and the second term is analogous
to the interaction (Coulomb repulsion) energy.

It is important that the Hamiltonian (\ref{IDS.1.6}) -- (\ref{IDS.1.7})
contains the factor $1/n$ in front of the second sum (interaction). This
allows us to view (\ref{IDS.1.6}) -- (\ref{IDS.1.7}) as an analog of
molecular field models of statistical mechanics. This analogy was implicitly
used in physical papers \cite{Wi:57,Dy:62,BIPZ:78}. A rigorous treatment of
a rather general class of mean field models was given in \cite%
{Pa-Sh:84,Sh:89}. We will use an extension of the treatment to study the
limit of NCM (\ref{NCM}), corresponding to (\ref{psymb}) -- (\ref{Delta}).
We stress a difference of this problem comparing to that of statistical
mechanics. In the latter the number of particles is explicitly present only
in the Hamiltonian (see the factor $1/n$ in the second term of (\ref{IDS.1.7}%
)), while in the former we have $n$ also in the Gibbs density (\ref{IDS.1.6}%
). In statistical mechanics terms we have here a mean field model in which
the temperature is inverse proportional to the number of particles, while in
a standard statistical mechanics treatment the temperature is fixed during
the "macroscopic limit" $n\rightarrow \infty $. This will imply that the
free energy of the model has to be divided by $n^{2}$ to have a well defined
limit as $n\rightarrow \infty $ and that the limit will coincide with the
limit as $n\rightarrow \infty $ of the ground state energy, also divided by $%
n^{2}$ (see \cite{BPS:95, Ki-Sp:99} and formulas (\ref{EV}) -- (\ref{infE}),
and (\ref{free_en}) below).

It is also well known in statistical mechanics that the macroscopic limit of
mean field models can be described in terms of certain extremal problems. In
our case the problem consists in minimizing the functional
\begin{equation}
\mathcal{E}[m]=\int V(\lambda )m(d\lambda )+\int \log \frac{1}{|\lambda -\mu
|}m(d\lambda )m(d\mu )  \label{EV}
\end{equation}%
defined on the set of non-negative unit measures $\mathcal{M}_{1}(\mathbb{R}%
) $ (cf (\ref{IDS.1.7})).

The variational problem \ (\ref{EV}) goes back to Gauss and is called the
minimum energy problem in the external field $V$. The unit measure $N$
minimizing (\ref{EV}) is called the equilibrium measure in the external
field $V$ because of its evident electrostatic interpretation as the
equilibrium distribution of linear charges on the ideal conductor occupying
the axis $\mathbb{R}$ and confined by the external electric field of
potential $V$. We stress that the corresponding variational procedure
determine both the (compact) support $\sigma _{N}$ of the measure and its
form. This should be compared with the widely known variational problem of
the theory of logarithmic potential, where the external field is absent but
the support is given (see e.g. \cite{La:72}). The minimum energy problem in
the external field (\ref{EV}) arises in various domains of analysis and its
applications (see \cite{Sa-To:97} for a rather complete account of results
and references concerning the problem).

\smallskip Before to start the systematic exposition we will make notational
conventions that will be used everywhere below. First, the integrals without
limits will denote the integrals over the whole axis. Second, symbols $%
C,c,C_{1},\dots $ etc. will denote positive finite quantities that do not
depend on $n$ and spectral variables and whose value is not important in the
corresponding argument.

\subsection{Basic results and their proof}

\smallskip We will need certain properties of the variational problem (\ref%
{EV}), given in the following

\begin{proposition}
\label{p:enpr} Let $V:\mathbb{R}\to\mathbb{R}_+$ be a continuous function
satisfying (\ref{cond0}). Then:

(i) there exists one and only one measure $N\in\mathcal{M}_1(\mathbb{R})$
such that
\begin{equation}  \label{infE}
\inf_{m\in\mathcal{M}_1(\mathbb{R})} \mathcal{E}[m]=\mathcal{E}[N]>-\infty,
\end{equation}
and $N$ has the finite logarithmic energy
\begin{equation}
\mathcal{L}[N,N]:= \displaystyle\int\log \frac{1}{| \lambda-\mu |}%
N(d\lambda)N(d\mu ) <\infty;  \label{IDS.1.10}
\end{equation}

(ii) the support $\sigma_N$ of $N$ is compact;

(iii) a measure $N\in\mathcal{M}_1(\mathbb{R})$ is as above if and only if
the function
\begin{equation}
u(\lambda;N)=V(\lambda)+2\int\log\frac{1}{\vert\lambda-\mu\vert} N(d\mu)
\label{IDS.1.11}
\end{equation}
satisfies the following relations almost everywhere with respect to $N$ (in
fact except the set of zero capacity):
\begin{equation}  \label{condin}
\begin{array}{ll}
u(\lambda;N)= u_*, &
\end{array}%
\end{equation}
where
\begin{equation}  \label{u_*}
u_*=\inf_{\lambda\in\mathbb{R}}u(\lambda;N)>-\infty;
\end{equation}

(iv) if the potential $V$ satisfies the H\"{o}lder condition
\begin{equation}  \label{HoldV}
\vert V(\lambda_1)-V(\lambda_2)\vert\le
C(L_1)\vert\lambda_1-\lambda_2\vert^{\gamma} , \quad\vert\lambda_{1,2}\vert
\le L_1
\end{equation}
for some $\gamma>0$ and any $L_1<\infty$, then $u(\lambda;N)$ also satisfies
the H\"{o}lder condition with the same $\gamma$:
\begin{equation}  \label{Holdu}
\vert u(\lambda_1;N)-u(\lambda_2;N)\vert\le
C^{\prime}(L_1)\vert\lambda_1-\lambda_2\vert^{\gamma} ,
\quad\vert\lambda_{1,2}\vert \le L_1;
\end{equation}

(v) if $m$ is a finite signed measure of zero charge, $m(\mathbb{R})=0$, or
its support belongs to $[-1,1]$, then
\begin{equation}  \label{Lpos}
\mathcal{L}[m,m]\ge 0,
\end{equation}
where for any finite signed measures $m_{1,2}$ we denote
\begin{equation}  \label{Lm_1,m_2}
\mathcal{L}[m_1,m_2] = \displaystyle\int\log \frac{1}{| \lambda-\mu |}%
m_1(d\lambda)m_2(d\mu ),
\end{equation}
$\mathcal{L}[m,m]=0$ if and only if $m=0$, we have
\begin{equation}  \label{LSchw}
|\mathcal{L}[m_1,m_2]|^2\le\mathcal{L}[m_1,m_1]\mathcal{L}[m_2,m_2],
\end{equation}
and (\ref{Lm_1,m_2}) defines a Hilbert structure on the space of signed
measures with a scalar product (\ref{Lm_1,m_2}) in which the convex cone of
non negative measures such that $\mathcal{L}[m,m]<\infty$ is complete, i.e.,
if $\{m^{(k)}\}_{k=1}^\infty$ is a sequence of non negative measures,
satisfying the Cauchy condition with respect to the norm (\ref{Lpos}), then
there exists a finite non-negative measure $m$ such that $m^{(k)}\to m$ in
the norm (\ref{Lpos}) and weakly;

(vi) if $m_{1,2}$ are finite signed measures with compact supports, and $%
m_{1}(\mathbb{R})=0$, then
\begin{equation}
\mathcal{L}[m_{1},m_{2}]=\int_{0}^{\infty }\frac{\widehat{m}_{1}(p)\widehat{m%
}_{2}(-p)}{p}dp,\quad \widehat{m}_{1,2}(p)=\int e^{ip\lambda
}m_{1,2}(d\lambda ).  \label{reprL}
\end{equation}
\end{proposition}

\textbf{Proof.} Assertions (i) -- (iii) are proved in Theorem I.1.3 and
I.3.3 of \cite{Sa-To:97} for not necessary continuous $V$, but it is shown
there only that $u(\lambda ;N)$ satisfies (\ref{condin}) almost everywhere
with respect to $N$. We will prove now that if $V$ is continuous, then $%
u(\lambda ;N)$ satisfies condition (\ref{condin}) for all $\lambda \in
\sigma _{N}$. To this end consider a point $\lambda _{0}\in \mathbb{R}$ such
that
\begin{equation*}
u(\lambda _{0};N)>u_{\ast }+\varepsilon ,\quad \varepsilon >0.
\end{equation*}%
Since $V$ is continuous, there exists $\delta _{1}>0$ such that
\begin{equation*}
V(\lambda )-V(\lambda _{0})>-\varepsilon /3,\quad |\lambda -\lambda
_{0}|\leq \delta _{1}.
\end{equation*}%
On the other hand, it is known \cite{La:72} that for any finite positive
measure $m$ the function
\begin{equation}
\mathcal{L}(\lambda ;m)=\int \log \frac{1}{|\lambda -\mu |}m(d\mu )
\label{Llm}
\end{equation}%
is upper semicontinuous, i.e. if $\mathcal{L}(\lambda _{0};m)<\infty $, then
for any $\varepsilon >0$ there exists $\delta _{2}>0$ such that
\begin{equation*}
\mathcal{L}(\lambda ;m)>\mathcal{L}(\lambda _{0};m)-\varepsilon /3,\quad
|\lambda -\lambda _{0}|\leq \delta _{2}.
\end{equation*}%
Using this property for $m=N$ we obtain from the above inequalities that
\begin{equation*}
u(\lambda )>u_{\ast }+\varepsilon /3,\quad |\lambda -\lambda _{0}|\leq
\delta :=\min \{\delta _{1},\delta _{2}\}.
\end{equation*}%
Then (\ref{condin}) and (\ref{u_*}) imply that $N((\lambda _{0}-\delta
,\lambda _{0}+\delta ))=0$, i.e., $\lambda _{0}\not\in \sigma _{N}$. For the
case $\mathcal{L}(\lambda _{0};N)=\infty $ the proof is the same.

Let us prove assertion (iv) of proposition. It is evident that it suffices
to prove that $\mathcal{L}(\lambda ;N)$ of (\ref{Llm}) is a H\"{o}lder
function in $\lambda $. If $\lambda _{1},\lambda _{2}\in \sigma _{N}$, then,
according to the above $2\mathcal{L}(\lambda _{1,2};N)=-V(\lambda
_{1,2})+u_{\ast }$, and (\ref{Holdu}) follows immediately from (\ref{HoldV}).

Since $\sigma _{N}$ is compact, $\mathbb{R}\setminus \sigma _{N}$ consists
of a finite or countable system of open intervals (gaps). Assume that $%
\lambda _{1},\lambda _{2}$ belong to the same gap $(\lambda _{1}^{\ast
},\lambda _{2}^{\ast })$: $\lambda _{1}^{\ast }<\lambda _{1}<\lambda
_{2}<\lambda _{2}^{\ast }$. Since $\mathcal{L}^{\prime \prime }(\lambda
;N)>0 $, $\lambda \in (\lambda _{1}^{\ast },\lambda _{2}^{\ast })$, and $%
\mathcal{L}(\lambda ;N)\geq (u_{\ast }-V(\lambda ))/2$, we have
\begin{eqnarray}
\frac{1}{2}\left( V(\lambda _{1}^{\ast })-V(\lambda _{1}^{\ast }+(\lambda
_{2}-\lambda _{1}))\right)& \leq& \mathcal{L}(\lambda _{1}^{\ast }+(\lambda
_{2}-\lambda _{1});N)-\mathcal{L}(\lambda _{1}^{\ast };N)  \label{Holdu2} \\
&\leq& \mathcal{L}(\lambda _{2};N)-\mathcal{L} (\lambda _{1};N)\leq \mathcal{%
L}(\lambda _{2}^{\ast };N)-\mathcal{L}(\lambda _{2}^{\ast }-(\lambda
_{2}-\lambda _{1});N)  \notag \\
&\leq& \frac{1}{2}\left( V(\lambda _{2}^{\ast }-(\lambda _{2}-\lambda
_{1}))-V(\lambda _{2}^{\ast })\right) ,  \notag
\end{eqnarray}%
and (\ref{Holdu}) follows from (\ref{HoldV}). Observe now, that this
inequality is also valid if $\lambda _{1}^{\ast }=\lambda _{1}$ or $\lambda
_{2}=\lambda _{2}^{\ast }$. The case when $\lambda _{1}$ or $\lambda _{2}$
belongs to semi infinite gap can be studied similarly.

If $\lambda _{1},\lambda _{2}$ belong to different gaps $\lambda _{1}\in
(\lambda _{1}^{\ast },\lambda _{2}^{\ast })$, $\lambda _{2}\in (\lambda
_{3}^{\ast },\lambda _{4}^{\ast })$, then (\ref{Holdu2}) implies
\begin{multline*}
|\mathcal{L}(\lambda _{1};N)-\mathcal{L}(\lambda _{2};N)|\leq |\mathcal{L}%
(\lambda _{1};N)-\mathcal{L}(\lambda _{2}^{\ast };N)|+|\mathcal{L}(\lambda
_{2}^{\ast };N)-\mathcal{L}(\lambda _{3}^{\ast };N)| \\
+|\mathcal{L}(\lambda _{3}^{\ast };N)-\mathcal{L}(\lambda _{2};N)|\leq
C(|\lambda _{1}-\lambda _{2}^{\ast }|^{\gamma }+|\lambda _{2}^{\ast
}-\lambda _{3}^{\ast }|^{\gamma }+|\lambda _{3}^{\ast }-\lambda
_{2}|^{\gamma }\leq 3^{1-\gamma }C|\lambda _{1}-\lambda _{2}|^{\gamma }.
\end{multline*}%
This proves assertion (iv).

Assertion (v) is proved in Theorem 1.16 of \cite{La:72}. Assertion (vi) is
proved in Lemma 6.41 of \cite{De:99} for the case $m_{2}({R})=0$. This
implies (\ref{reprL}) for a general case of $m_{2}$. The proposition is
proved.

\medskip

We formulate now the main result of this section.

\begin{theorem}
\label{t:IDS.t1} Consider a collection of random variables $\{\lambda
_{l}^{(n)}\}_{l=1}^{n}$, specified by the probability density (\ref{psymb})
-- (\ref{Delta}) in which $\beta >0$ and the potential $V$ satisfies (\ref%
{cond0}) and (\ref{HoldV}). Then:

(i) there exists $0<L<\infty$ such that for any $|\lambda_1|,|\lambda_2|\ge
L $
\begin{equation}
\rho_n(\lambda_1)\le e^{-nC V(\lambda_1)},\quad
p^{(n)}_{2,\beta}(\lambda_1,\lambda_2)\le e^{-nC(
V(\lambda_1)+V(\lambda_2))},  \label{exp_est}
\end{equation}
where $\rho_n$ and $p^{(n)}_{2,\beta}$ are defined in (\ref{Nnrho}) and (\ref%
{pnlb}), and $L$ depends on $\epsilon$ of (\ref{cond0}) and on
\begin{equation}
m=\min_{\lambda\in \mathbb{R}}\{V(\lambda)-2(1+\epsilon/2)\log(1+|\lambda|)%
\},\quad M=\max_{\lambda\in[-1/2,1/2]}V(\lambda);  \label{M}
\end{equation}

(ii) the Normalized Counting Measure (\ref{NCM}) of the collection $%
\{\lambda _{l}^{(n)}\}_{l=1}^{n}$ converges in probability to the unique
minimizer $N$ of (\ref{EV}) -- (\ref{infE}), and for any differentiable
function $\varphi :[-L,L]\rightarrow \mathbb{C}$ we have
\begin{equation}
\left\vert \int \varphi (\mu )\rho _{n}(\mu )d\mu -\int \varphi (\mu )N(d\mu
)\right\vert \leq C||\varphi ^{\prime }||_{2}^{1/2}||\varphi
||_{2}^{1/2}\cdot n^{-1/2}\log ^{1/2}n,  \label{w_conv}
\end{equation}%
\begin{equation}
\left\vert \int \varphi (\lambda )\varphi (\mu )(p_{2}^{(n)}(\lambda ,\mu
)-\rho _{n}(\lambda )\rho _{n}(\mu ))d\lambda d\mu \right\vert \leq
C||\varphi ^{\prime }||_{2}||\varphi ||_{2}\cdot n^{-1}\log n,  \label{fact}
\end{equation}%
where the symbol $||...||_{2}$ denotes the $L^{2}$-- norm on $[-L,L]$;

(iii) the free energy $-2(\beta n^{2})^{-1}\log Q_{n,\beta }$ of the model (%
\ref{IDS.1.6}) -- (\ref{Qnb}) converges as $n\rightarrow \infty $ to the
ground state energy (\ref{infE}) and
\begin{equation}
\bigg|\frac{2}{\beta n^{2}}\log Q_{n,\beta }+\mathcal{E}[N]\bigg|\leq
Cn^{-1}\log n.  \label{free_en}
\end{equation}
\end{theorem}

\begin{theorem}
\label{t:IDS.2} Let $V$ satisfy (\ref{cond0}) and $V^{\prime }$ be such that
for any $A>0$ there exists $C(A)>0$ providing the inequality
\begin{equation}
|V^{\prime }(\lambda )-V^{\prime }(\mu )|\leq C(A)|\lambda -\mu |,\quad
|\lambda |,|\mu |\leq A.  \label{Lip}
\end{equation}%
Consider the measure $N$ defined by (\ref{infE}) and denote $f$ its
Stieltjes transform:
\begin{equation}
f(z)=\int \frac{N(d\lambda )}{\lambda -z},\quad \Im z\not=0.  \label{f}
\end{equation}%
Then $f$ satisfies the equation
\begin{equation}
f^{2}(z)=\int \frac{V^{\prime }(\lambda )N(d\lambda )}{\lambda -z},
\label{sq-eq}
\end{equation}%
$N$ has a bounded density $\rho $ which can be represented in the form
\begin{equation}
\rho (\lambda )=\frac{1}{2\pi }\left( 4Q(\lambda )-V^{\prime 2}(\lambda
)\right) _{+}^{1/2},  \label{sq-rep}
\end{equation}%
where $(x)_{+}=\max \{x,0\}$,
\begin{equation}
Q(\lambda )=\int \frac{V^{\prime }(\lambda )-V^{\prime }(\mu )}{\lambda -\mu
}\rho (\mu )d\mu ,  \label{Q}
\end{equation}%
and we have
\begin{equation}
|\rho ^{2}(\lambda )-\rho ^{2}(\mu )|\leq C|\lambda -\mu |\log \frac{1}{%
|\lambda -\mu |}.  \label{d-rho}
\end{equation}
\end{theorem}

\noindent The proof of the theorem is based on the ideas of \cite{Pa:96b}
(see also \cite{De-Co:98}) and is given below, after the proof of Theorem %
\ref{t:IDS.t1}.

\begin{remark}
\label{r:IDS.t2} It follows from the theorem that under condition (\ref{Lip}%
) we can differentiate the r.h.s. of (\ref{condin}) with respect to $\lambda$%
. Then we obtain that $\rho$ solves the singular integral equation
\begin{equation}  \label{sieq}
V^{\prime}(\lambda)=2\int_{\sigma}\frac{\rho(\mu)d\mu}{\lambda-\mu},\quad
\lambda\in\sigma.
\end{equation}
\end{remark}

\begin{theorem}
\label{t:IDS.3} Let $V$ satisfy conditions of Theorem \ref{t:IDS.2}
and $u(\lambda )\not=u_{\ast }$ for $\lambda \not\in \sigma _{N}$ (see (\ref%
{IDS.1.11}), (\ref{condin})). Denote by $\sigma _{N}^{(\varepsilon )}$ the $%
\varepsilon $-neighborhood of $\sigma _{N} $ and
\begin{equation}
d_{n}=\int_{\mathbb{R}\setminus \sigma _{N}}e^{-\beta n(u(\lambda )-u_{\ast
})/4}d\lambda ,\quad d(\varepsilon )=\sup_{\mathbb{\ R}\setminus \sigma
_{N}^{(\varepsilon )}}\{(u_{\ast }-u(\lambda ))/4\}  \label{e_n}
\end{equation}%
Then there exists an $n$-independent $C>0$ such that for any $\varepsilon >0$
(may be depending on $n$), satisfying condition $d(\varepsilon
)>C(n^{-1/2}\log n+d_{n})$ we have the bound (cf (\ref{exp_est}))
\begin{equation}
\overline{N}_{n}(\mathbb{R}\setminus \sigma _{N}^{(\varepsilon )})\leq
e^{-nd(\varepsilon )},  \label{tails}
\end{equation}%
where $\overline{N}_{n}$ is defined in (\ref{Nbp}).
\end{theorem}

\noindent The proof of the theorem is given below after the proof of Theorem %
\ref{t:IDS.2}.

\begin{remark}
\label{r:IDS.t3} It follows from the proof of the theorem that if we replace
(\ref{Lip}) by conditions (\ref{HoldV}) and $|\sigma _{N}|\not=0$, then
Theorem \ref{t:IDS.3} will also be valid.
\end{remark}

\begin{remark}
\label{r:IDS.t3a} Usually $d_n$ of (\ref{e_n}) is $O(n^{-1})$, but it may
happen also that $d_n\to 0$ vanishes more slowly as $n\to\infty$.
\end{remark}

\textbf{Proof of Theorem \ref{t:IDS.t1}.} Following the main idea of \cite%
{Pa-Sh:84,Sh:89} we will use the Bogolyubov inequality (a version of the
Jensen inequality) to control the free energy of our "mean field" model. The
inequality is given by

\begin{lemma}
\label{l:Bog} Let $\mathcal{H}_{1,2}:\mathbb{R}^n\to\mathbb{R}$ be such that
\begin{equation*}
Q_{1,2}:=\int e^{-\mathcal{H}_{1,2}(\Lambda)/T}d\Lambda<\infty,\quad
\Lambda=(\lambda_1,\dots,\lambda_n)\in\mathbb{R}^n,\quad T>0.
\end{equation*}
Denote
\begin{equation*}
\langle\dots\rangle_{1,2}=Q_{1,2}^{-1}\int\dots e^{-\mathcal{H}%
_{1,2}(\Lambda)/T}d\Lambda.
\end{equation*}
Then
\begin{equation}
\langle\mathcal{H}_1-\mathcal{H}_2\rangle_{1}\le T\log Q_{2}-T\log Q_{1} \le
\langle \mathcal{H}_1-\mathcal{H}_2\rangle_{2}.  \label{Bog}
\end{equation}
\end{lemma}

\noindent The proof of the lemma is given in the next subsection.

\smallskip Since the proof of assertion (i) is independent of the proof of
(central) assertion (ii), we will give the proof assertion (i) in the next
subsection. We will use however assertion (i) in the proof of assertion (ii).

According to assertion (i) the limiting measure $N$ of (\ref{Nbp}), if it
exists, has its support strictly inside $[-L,L]$. Let us show that the
limiting measure does not depend on values of the potential outside $[-L,L]$%
. To this end consider potentials $V_{1}$ and $V_{2}$, verifying conditions (%
\ref{cond0}) and (\ref{HoldV}). Then the potential%
\begin{equation}
V(\lambda ,t)=tV_{1}(\lambda )+(1-t)V_{2}(\lambda )  \label{Vtl}
\end{equation}%
also satisfies (\ref{cond0}) and (\ref{HoldV}). Denote $\overline{N}%
_{n}(\cdot ,t),\;\rho _{n}(\cdot ,t)$, and $p_{2,\beta }^{(n)}(\cdot ,\cdot
,t)$ the measure (\ref{Nnrho}), its density, and the second marginal of (\ref%
{psymb}) corresponding to (\ref{Vtl}). Then it is easy to find from (\ref%
{psymb}) -- (\ref{pnlb}) that%
\begin{eqnarray}
\frac{\partial }{\partial t}\rho _{n}(\lambda ,t) &=&-n\delta V(\lambda
)\rho _{n}(\lambda ,t)-n(n-1)\int \delta V(\mu )p_{2,\beta }^{(n)}(\lambda
,\mu ,t)d\mu  \label{dtro} \\
&&+n^{2}\rho _{n}(\lambda ,t)\int \delta V(\mu )\rho _{n}(\mu ,t)d\mu ,
\notag
\end{eqnarray}%
where $\delta V=V_{1}-V_{2}$. This implies the bound
\begin{equation*}
\left\vert \frac{\partial }{\partial t}\overline{N}_{n}(\Delta
,t)\right\vert \leq 2n^{2}\int |\delta V(\mu )|\rho _{n}(\mu ,t)d\mu ,
\end{equation*}%
valid for any $\Delta \in \mathbb{R}$. Now, if $V_{1}(\lambda
)=V_{2}(\lambda )$, $|\lambda |<L$, then in view of (\ref{exp_est}) and (\ref%
{cond0}) we have:
\begin{equation}
\begin{array}{c}
\bigg|\overline{N}_{n}(\Delta )\bigg|_{V=V_{1}}-\overline{N}_{n}(\Delta )%
\bigg|_{V=V_{2}}\bigg|\leq 2n^{2}\displaystyle\int_{|\lambda |>L}d\lambda
|V_{1}(\lambda )-V_{2}(\lambda )|\int_{0}^{1}e^{-nCV(\lambda ,t)}dt \\
\hspace{3cm}\leq 2C^{-1}n\displaystyle\int_{|\lambda
|>L}(e^{-nCV_{1}(\lambda )}+e^{-nCV_{2}(\lambda )})d\lambda
=O(e^{-nC^{\prime }}).%
\end{array}
\label{dNbo}
\end{equation}%
We conclude that without loss of generality we can assume that the potential
satisfies the H\"{o}lder condition on the whole axis with the same exponent
as in (\ref{HoldV}):%
\begin{equation}
|V(\lambda _{1})-V(\lambda _{2})|\leq C|\lambda _{1}-\lambda _{2}|^{\gamma
},\;\lambda _{1},\lambda _{2}\in \mathbb{R}.  \label{HoldVev}
\end{equation}%
Furthermore, we can also assume without loss of generality that the
parameter $L$ of assertion (i) of the theorem is less than $1/2$ and that
the support $\sigma _{N}$ of the minimizer $N$ of (\ref{EV}) -- (\ref{infE})
and all the points $\lambda _{k}^{\ast }$ such that $u(\lambda _{k}^{\ast
})=u_{\ast }$ are contained in the interval $[-1/2+\delta ,1/2-\delta ]$ for
some $\delta >0$.

Let us prove (\ref{w_conv}). Denote by $\mathcal{C}^{\ast }$ the cone of
measures on $\mathbb{R}$ satisfying the conditions:
\begin{equation}
m(d\lambda )\geq 0,\;\;\hbox{supp}\,\,m\subset \lbrack -1/2,1/2],\quad
\mathcal{L}[m,m]<\infty ,\quad m(\mathbb{R} )\leq 1.  \label{cond_m}
\end{equation}%
For any $m\in \mathcal{C}^{\ast }$ we introduce the "approximating"
Hamiltonian
\begin{equation}  \label{IDS.2.8}
H_{a}(\Lambda ;m)=\displaystyle\sum_{i=1}^{n}u_n(\lambda_i;m)-(n-1)\mathcal{L%
}[m,m],
\end{equation}%
where (cf (\ref{IDS.1.11}))
\begin{equation}  \label{u_n}
u_{n}(\lambda ;m)=V(\lambda )+2\frac{n-1}{n}\mathcal{L}(\lambda ;m),
\end{equation}%
and $\mathcal{L}(\lambda ;m)$, $\mathcal{L}[m,m]$ are defined by (\ref{Llm})
and (\ref{Lm_1,m_2}). Consider the functional $\Phi_n:\mathcal{C}^*\to%
\mathbb{R}$, defined as
\begin{equation}  \label{IDS.2.9}
\Phi _{n}[m] =\frac{2}{\beta n^{2}}\log \int e^{-\beta nH_{a}(\Lambda
;m)/2}d\Lambda =\frac{(n-1)}{n}\mathcal{L}[m,m]+\frac{2}{\beta n}\log \int
e^{-\beta n u_n(\lambda;m)/2} d\lambda .
\end{equation}%
Taking in (\ref{Bog}) $\mathcal{H}_{1}=H$, $\mathcal{H}_{2}=H_{a}$ and $%
T=2/\beta n$, we obtain
\begin{equation}
R[m]\leq \Phi _{n}[m]-\frac{2}{\beta n^{2}}\log Q_{n,\beta }\leq R_{a}[m],
\label{IDS.2.10}
\end{equation}%
where
\begin{equation*}
\begin{array}{rcl}
R[m] & = & 2(\beta n^{2}Q_{n,\beta })^{-1}\displaystyle\int (H-H_{a})
e^{-\beta nH/2}d\Lambda , \\
R_{a}[m] & = & 2(\beta n^{2})^{-1}e^{-\beta n^{2}\Phi _{n}[m]/2}\displaystyle%
\int (H-H_{a})e^{-\beta n H_{a}(\Lambda ;m)/2}d\Lambda ,%
\end{array}%
\end{equation*}%
and $Q_{n,\beta }$ is defined in (\ref{Qnb}). Since $H$ and $H_{a}$ are
symmetric, we can rewrite $R[m]$ as follows
\begin{equation}
R[m]=\frac{n-1}{n}\left(\int \log \displaystyle\frac{1}{|\lambda -\mu |}%
(p_{2,\beta }^{(n)}(\lambda ,\mu )-\rho _{n}(\lambda )\rho _{n}(\mu
))d\lambda d\mu +\mathcal{L}[\overline{N}_{n}-m,\overline{N}_{n}-m]\right),
\label{IDS.R}
\end{equation}%
where $p_{2,\beta }^{(n)}$, $\rho _{n}$, and $\overline{N}_{n}$ are defined
in (\ref{pnlb}), (\ref{Nbp}) -- (\ref{Nnrho}). To obtain $R_{a}$, we have to
replace $\rho _{n}(\lambda )$ and $p_{2,\beta }^{(n)}(\lambda ,\mu )$ in (%
\ref{IDS.R}) by $\rho _{n}^{(a)}(\lambda ;m)$ and $\rho _{n}^{(a)}(\lambda
;m)\rho _{n}^{(a)}(\mu ;m)$, the correlation functions of the ~approximating
Hamiltonian (\ref{IDS.2.8}), where
\begin{equation}  \label{rho_a}
\rho _{n}^{(a)}(\lambda ;m)=e^{-\beta nu_n(\lambda;m )/2} \bigg(\int
d\lambda e^{-\beta nu_n(\lambda;m )/2}\bigg)^{-1}.
\end{equation}%
This yields:
\begin{equation}
R_{a}[m]=\frac{n-1}{n}\mathcal{L}[N_{n}^{(a)}-m,N_{n}^{(a)}-m],  \label{Ram}
\end{equation}%
where
\begin{equation}
N_{n}^{(a)}(d\lambda ;m)=\rho _{n}^{(a)}(\lambda ;m)d\lambda .  \label{Nan}
\end{equation}

\begin{lemma}
\label{l:IDS.l2} Let $\mathcal{C}^{\ast }$ be the cone of measures defined by (%
\ref{cond_m}) and the functional $\Phi _{n}:\mathcal{C}^{\ast }\rightarrow
\mathbb{R}$, be given by (\ref{IDS.2.9}). Then,

(i) $\Phi _{n}$ attains its minimum at a unique point $m_n\in \mathcal{C}%
^{\ast }$  and
\begin{equation}
\mathcal{L}[N_{n}^{(a)}-m_n,N_{n}^{(a)}-m_n]\leq e^{-nc};
\label{exp_est4}
\end{equation}

(ii) if $N$ is a measure, defined by (\ref{IDS.1.7}) -- (\ref{IDS.1.11}),
then
\begin{equation}
0\leq \Phi _{n}[N]-\Phi _{n}[m_n]\leq Cn^{-1}\log n.  \label{est}
\end{equation}
\end{lemma}

\noindent The proof of Lemma \ref{l:IDS.l2} is given in the next subsection.

\smallskip

On the basis of (\ref{IDS.2.10}), Lemma \ref{l:IDS.l2}, and (\ref{Ram}) we
obtain
\begin{eqnarray}
R[N] &\leq &\Phi _{n}[N]-\frac{2}{\beta n^{2}}\log Q_{n,\beta }  \label{b_1}
\\
&=&(\Phi _{n}[N]-\Phi _{n}[m_n])+(\Phi _{n}[m_n]-\frac{2}{\beta n^{2}}\log
Q_{n,\beta })  \notag \\
&\leq &Cn^{-1}\log n+R_{a}[m_n]\leq Cn^{-1}\log n+Ce^{-nc}.  \notag
\end{eqnarray}%
This and (\ref{IDS.R}) lead to the inequality%
\begin{eqnarray}
&&\int \log \frac{1}{|\lambda -\mu |}(p_{2}^{(n)}(\lambda ,\mu )-\rho
_{n}(\lambda )\rho _{n}(\mu ))d\lambda d\mu  \label{b_2} \\
&&\hspace{2cm}+\mathcal{L}[\overline{N}_{n}-N,\overline{N}_{n}-N])\leq
Cn^{-1}\log n.  \notag
\end{eqnarray}%
Since $\mathcal{L}[\overline{N}_{n}-N,\overline{N}_{n}-N]\geq 0$ by
Proposition \ref{p:enpr} (v), we have the bound
\begin{equation}
\int \log \frac{1}{|\lambda -\mu |}G_{n}(\lambda ,\mu )d\lambda d\mu \leq C%
\frac{\log n}{n},\quad G_{n}(\lambda ,\mu )=p_{2}^{(n)}(\lambda ,\mu )-\rho
_{n}(\lambda )\rho _{n}(\mu ).  \label{Gnl}
\end{equation}%
We will prove now that there exists an $n$-independent $C>0$ such that
\begin{equation}
\int \log \frac{1}{|\lambda -\mu |}G_{n}(\lambda ,\mu )d\lambda d\mu ,\geq -C%
\frac{\log n}{n},  \label{b_Gb}
\end{equation}%
and, as a result, that
\begin{equation}
\int \log \frac{1}{|\lambda -\mu |}G_{n}(\lambda ,\mu )d\lambda d\mu
=O(n^{-1}\log n).  \label{b_G}
\end{equation}%
Note that (\ref{b_G}) and (\ref{b_1}) yield assertion (iii) of Theorem \ref%
{t:IDS.t1}. Indeed, it follows from (\ref{b_2}) and (\ref{b_G}) that
\begin{equation}
\mathcal{L}[\overline{N}_{n}-N,\overline{N}_{n}-N]=O(n^{-1}\log n).
\label{b_L}
\end{equation}%
This and (\ref{b_1}) imply
\begin{equation}
\Phi _{n}[N]-\frac{2}{\beta n^{2}}\log Q_{n,\beta }=O(n^{-1}\log n).
\label{free_en.1}
\end{equation}%
Since according to (\ref{Holdu}) $\mathcal{L}(\lambda ;N)$ is a H\"{o}lder
function, it is easy to find by the Laplace method that
\begin{eqnarray*}
\Phi _{n}[N] &=&\frac{n-1}{n}\mathcal{L}[N;N]-\min_{\lambda }\{u(\lambda
;N)\}+O(n^{-1}\log n) \\
&=&\frac{n-1}{n}\mathcal{L}[N;N]-\int u(\lambda ;N)N(d\lambda )+O(n^{-1}\log
n) \\
&=&-\mathcal{E}[N]+O(n^{-1}\log n).
\end{eqnarray*}%
Here $u(\lambda ;N)$ is defined by (\ref{IDS.1.11}) and we have used (\ref%
{condin}). The two last relations yield (\ref{free_en}).

To prove (\ref{b_Gb}) we need certain upper bounds for $\rho _{n}$ and $%
p_{2}^{(n)}$. Changing variables $\lambda_i\to\lambda_i-x$ and using (\ref%
{HoldVev}) we find that for any $|x|\le h:=n^{-3/\gamma }$
\begin{eqnarray}
&&\bigg|\rho _{n}(\lambda _{1}+x)-\rho _{n}(\lambda _{1})\bigg|= Q_{n,\beta
}^{-1}\bigg|\int d\lambda _{2}...d\lambda _{n}\cdot |\Delta (\Lambda
)|^{\beta }  \label{d_rho} \\
&&\times e^{-nV(\lambda _{1}+x)}\prod_{i=2}^{n}e^{-nV(\lambda _{i}-x)}
-\prod_{i=2}^{n}e^{-nV(\lambda _{i})}\bigg|\leq \,Cn^2x^{\gamma }\rho
_{n}(\lambda _{1}).  \notag
\end{eqnarray}
Now we use the simple identity valid for any interval $[a,b]$ and any
integrable function $v(\lambda)$
\begin{equation}  \label{triv_id}
v(\lambda)=(b-a)^{-1}\int_a^b\left(v(\lambda)-v(\mu)\right)d\mu+
(b-a)^{-1}\int_a^bv(\mu)d\mu.
\end{equation}%
The identity with $v(\lambda)=\rho_n(\lambda)$, $a=\lambda$, $b=\lambda+h$, (%
\ref{d_rho}), and the normalization condition
\begin{equation}
\int \rho _{n}(\lambda )d\lambda =1  \label{inrho}
\end{equation}%
lead to the inequality
\begin{equation*}
\rho _{n}(\lambda )\leq Cn^{-1}\rho_n(\lambda)+n^{3/\gamma},
\end{equation*}
implying
\begin{equation}
\rho _{n}(\lambda )\leq C n^{3/\gamma}.  \label{b_rho.1}
\end{equation}%
Similarly we have for $p_{2}^{(n)}$ of (\ref{pnlb}), and $G_{n}$ of (\ref%
{Gnl}):
\begin{equation}
p_{2}^{(n)}(\lambda ,\mu )\leq \,C\,n^{6/\gamma },\quad \int
G_{n}^{2}(\lambda ,\mu )d\lambda d\mu \leq \,C\,n^{6/\gamma }.
\label{b_rho.2}
\end{equation}%
Furthermore, we can write the equality
\begin{equation}
\log |t|^{-1}=\sum_{k=-\infty }^{\infty }l^{(k)}e^{ikt\pi },\quad |t|\leq 1,
\label{Four_l}
\end{equation}%
valid in $L^{2}([-1,1])$ and in which
\begin{eqnarray}
&&\frac{C_{2}}{|k|}<l^{(k)}=\frac{1}{\pi |k|}\int_{0}^{\pi |k|}\frac{\sin x}{%
x}dx<\frac{C_{1}}{|k|},\;k\not=0,  \label{l_k}
\end{eqnarray}%
Besides, since for any bounded continuous function $f:\mathbb{R}\rightarrow
\mathbb{C}$ we have
\begin{equation*}
\int \bigg|\frac{1}{n}\sum_{i=1}^{n}(f(\lambda _{i})-\langle f\rangle )\bigg|%
^{2}p_{n}(\Lambda )d\Lambda \geq 0,\quad \langle f\rangle =\int f(\lambda
)\rho _{n}(\lambda )d\lambda ,
\end{equation*}%
the symmetry of $p_{n}$ of (\ref{psymb}) implies:
\begin{equation}
\int f(\lambda )\overline{f}(\mu )G_{n}(\lambda ,\mu )d\lambda d\mu
+(n-1)^{-1}\langle |f|^{2}\rangle \geq 0.  \label{IDS.t1.4}
\end{equation}%
We now write integral in (\ref{b_Gb}) as that over the square $%
\{|\lambda|\le 1/2,|\mu|\le 1/2\}$ and over the complement of the square.
The second integral is $O(e^{-nc})$ by (\ref{exp_est}) and (\ref{b_rho.2}).
In the first integral we replace $\log|\lambda-\mu|^{-1}$ by the r.h.s. of (%
\ref{Four_l}) with $t=\lambda-\mu$. Thus, choosing $M=n^{2+6/\gamma }$, we
get:
\begin{eqnarray}
&&\int \log \frac{1}{|\lambda -\mu |}G_{n}(\lambda ,\mu )d\lambda d\mu
=O(e^{-nc})+\sum_{k\not=0}G_{n}^{(k,k)}l^{(k)}  \label{IDS.t1.2} \\
&&\hspace{1cm}=\sum_{k<M}(G_{n}^{(k,k)}+(n-1)^{-1})l^{(k)}-(n-1)^{-1}%
\sum_{k<M}l^{(k)}+R_{M}  \notag \\
&&\hspace{1cm}\geq -Cn^{-1}\log n+R_{M}.  \notag
\end{eqnarray}%
Here
\begin{equation*}
G_{n}^{(k,m)}=\int_{-1/2}^{1/2}e^{ik\pi \lambda -im\pi \mu }G_{n}(\lambda
,\mu )d\lambda d\mu ,
\end{equation*}%
and we use (\ref{exp_est}) implying $G_{n}^{(0,0)}=O(e^{-Cn})$ and (\ref%
{IDS.t1.4}) implying $G_{n}^{(k,k)}+n^{-1}\geq 0$. To estimate $R_{M}$ we
use (\ref{l_k}) and (\ref{b_rho.2}) to write
\begin{equation}
|R_{M}|\leq \sum_{k>M}|G_{n}^{(k,k)}l^{(k)}|\leq \bigg[%
\sum_{k,m}|G_{n}^{(k,m)}|^{2}\bigg]^{1/2}\bigg[\sum_{k>M}|l^{(k)}|^{2}\bigg]%
^{1/2}\leq C\frac{n^{3/\gamma }}{M^{1/2}}.  \label{IDS.t1.3}
\end{equation}%
The bound (\ref{b_G}) follows from (\ref{IDS.t1.2}) and (\ref{IDS.t1.3}).

Consider now a function $\varphi :[-1/2,1/2]\rightarrow \mathbb{C}$ such
that $\varphi ^{\prime }\in L_{2}[-1/2,1/2]$ and denote
\begin{equation*}
\begin{array}{l}
\varphi _{1}(\lambda )=\varphi (\lambda )\mathbf{1}_{|\lambda |\leq
1/2}+2\varphi (-1/2)(1+\lambda )\mathbf{1}_{\lambda <-1/2}+2\varphi
(1/2)(1-\lambda )\mathbf{1}_{\lambda >1/2},\quad |\lambda|\le 1, \\
\varphi ^{(k)}=\displaystyle\frac{1}{2}\int_{-1}^{1}\varphi _{1}(\lambda
)e^{ik\pi \lambda }d\lambda , \\
d^{(k)}=\displaystyle\frac{1}{2}\int_{-1}^{1}e^{ik\pi \lambda }(N(d\lambda )-%
\overline{N}_{n}(d\lambda )).%
\end{array}%
\end{equation*}%
Then, using (\ref{exp_est}), (\ref{l_k}), and the Parseval equation, we get
\begin{equation*}
\begin{array}{l}
\bigg|\int \varphi (\lambda )(N(d\lambda )-\overline{N}_{n}(d\lambda )) %
\bigg|^{2}=\bigg|O(e^{-nc})+\displaystyle\sum_{k\in \mathbb{Z}}\varphi
^{(k)}d^{(k)}\bigg|^{2} \\
\leq C\displaystyle\sum_{k\in \mathbb{Z}}l^{(k)}|d^{(k)}|^{2}\displaystyle%
\sum_{k\in \mathbb{Z}}|k||\varphi ^{(k)}|^{2}+O(e^{-nc})\leq C\mathcal{L}[N-%
\overline{N}_{n},N-\overline{N}_{n}]\cdot ||\varphi ||_{2}||\varphi ^{\prime
}||_{2}+O(e^{-nc}).%
\end{array}%
\end{equation*}
This inequality and (\ref{b_L}) imply (\ref{w_conv}). The inequality (\ref%
{fact}) can be proved similarly.

We will prove now that for any finite interval $\Delta\subset\mathbb{R}$ $%
N_n(\Delta)$ of (\ref{NCM}) converges in probability to $N(\Delta)$ of (\ref%
{infE}) as $n\to\infty$, i.e. that for any $\varepsilon>0$
\begin{equation*}
\lim_{n\to\infty}\mathbf{P}\{|N(\Delta)-N_n(\Delta)|>\varepsilon\}=0,
\end{equation*}
where $\mathbf{P}\{...\}$ denotes the probability, corresponding to (\ref%
{psymb}). This is the first part of assertion (ii) of Theorem \ref{t:IDS.t1}.

Denote $\Delta =(a,b)$, $-\infty <a<b<\infty $, $\chi $ the indicator of $%
\Delta $, and $\chi _{+}$ the continuous function, coinciding with $\chi $
on $(a,b)$, equal zero outside $(a-\delta ,b+\delta )$ for a sufficiently
small $\delta $ and linear on $(a-\delta ,a)$ and $(b,b+\delta )$. Let $\chi
_{-}$ be the analogous function for the interval $(a+\delta ,b-\delta )$.
Then
\begin{equation}
\chi _{-}\leq \chi \leq \chi _{+},\quad ||\chi _{\pm }||_{2}^{2}\leq
b-a+\delta ,\quad ||\chi _{\pm }^{\prime }||_{2}^{2}\leq 2\delta ^{-1}.
\label{chi}
\end{equation}%
Hence
\begin{equation}
N_{n}[\chi _{-}]\leq N_{n}[\chi ]\leq N_{n}[\chi _{+}],  \label{Nnpm}
\end{equation}%
where we denote for any $\varphi :\mathbb{R}\rightarrow \mathbb{C}$
\begin{equation}
N_{n}[\varphi ]=n^{-1}\sum_{l=1}^{n}\varphi (\lambda _{l}^{(n)})=\int
\varphi (\lambda )N_{n}(d\lambda ).  \label{linst}
\end{equation}%
This is a normalized linear statistics of random variables $\{\lambda
_{l}^{(n)}\}_{l=1}^{n}$, corresponding to the test function $\varphi $. We
have in particular $N_{n}[\chi ]=N_{n}(\Delta )$. By using this notation, we
can rewrite (\ref{w_conv}) as
\begin{equation}
|\mathbf{E}\{N_{n}[\varphi ]\}-N[\varphi ]|\leq Cn^{-1/2}\log
^{1/2}n||\varphi ||_{2}^{1/2}||\varphi ^{\prime }||_{2}^{1/2},
\label{w_conv.1}
\end{equation}%
where $\mathbf{E}\{...\}$ denotes the expectation with respect to (\ref%
{psymb}) and
\begin{equation*}
N[\varphi ]=\int \varphi (\lambda )N(d\lambda ).
\end{equation*}%
Choosing in (\ref{w_conv.1}) $\varphi =\chi _{\pm }$, taking into account (%
\ref{chi}) and the continuity of $N$ and making first the limit $%
n\rightarrow \infty $ and then $\delta \rightarrow 0$, we obtain
\begin{equation}
\lim_{n\rightarrow \infty }\mathbf{E}\{N_{n}(\Delta )\}=N(\Delta ).
\label{limE}
\end{equation}%
Likewise, denoting
\begin{equation*}
\mathbf{Var}\{N_{n}[\varphi ]\}=\mathbf{E}\{N_{n}^{2}[\varphi ]\}-\mathbf{E}%
^{2}\{N_{n}[\varphi ]\},
\end{equation*}%
we obtain from (\ref{fact})
\begin{equation}
\mathbf{Var}\{N_{n}[\varphi ]\}\leq Cn^{-1}\log n\,\,||\varphi
||_{2}||\varphi ^{\prime }||_{2}.  \label{Var_est}
\end{equation}%
Using this bound, (\ref{chi}) and (\ref{Nnpm}) we obtain
\begin{equation}
\lim_{n\rightarrow \infty }\mathbf{Var}\{N_{n}(\Delta )\}=0.  \label{limV}
\end{equation}%
Formulas (\ref{limE}) and (\ref{limV}) imply the convergence of the sequence
$\{N_{n}[\Delta ]\}$ in probability to the non random limit $N(\Delta )$ for
any finite $\Delta \subset \mathbb{R}$. Theorem \ref{t:IDS.t1} is proved.

\begin{remark}
\label{r:IDS.t1} Inspecting the above proof of Theorem \ref{t:IDS.t1}, we
conclude that its assertions remain valid if we replace the potential $V$ in
(\ref{psymb}) by $V+\varepsilon _{n}V_{1}$, where $V_{1}$ satisfies (\ref%
{cond0}) and (\ref{HoldV}) and $\varepsilon _{n}=O(n^{-1}\log n)$. If $%
\varepsilon _{n}\rightarrow 0$ more slowly, than $n^{-1/2}\log ^{-1/2}n$ and
$n^{-1}\log ^{-1}n$ in the r.h.s. of (\ref{w_conv}) and (\ref{fact}) should
be replaced by $\varepsilon _{n}^{1/2}$ and $\varepsilon _{n}$ respectively.
\end{remark}

\textbf{Proof of Theorem \ref{t:IDS.2}.} We follow the idea of \cite{Pa:96b}
(see also \cite{De-Co:98}). Consider a collection of random variables $%
\{\lambda _{l}^{(n)}\}_{l=1}^{n}$, specified by the probability density (\ref%
{psymb}) -- (\ref{Delta}) for $\beta =2$. We remark first that without loss
of generality we can assume that $V(\lambda )$ is a linear function outside
of the interval $[-L,L]$, where $L$ is defined in assertion (i) of Theorem %
\ref{t:IDS.t1} and hence, in view of (\ref{Lip}), that
\begin{equation}
\sup_{\lambda \in \mathbb{R}}|V^{\prime }(\lambda )|\leq C<\infty .
\label{b_V'}
\end{equation}%
Indeed, it suffices to repeat the argument, leading to (\ref{HoldVev}).

We have from (\ref{psymb}) -- (\ref{pnlb}) for $\beta=2$ and $l=1$:
\begin{equation}
\begin{array}{lll}
\rho _{n}(\lambda ) & = & Q_{n,2}^{-1}\displaystyle\int e^{-nV(\lambda )}%
\displaystyle\prod_{j=2}^{n}d\lambda _{j}e^{-nV(\lambda _{j})}(\lambda
-\lambda _{j})^{2}\displaystyle\prod_{2\leq j<k\leq n}(\lambda _{j}-\lambda
_{k})^{2}.%
\end{array}
\label{p2.1}
\end{equation}%
Then, taking any $z$ with $\Im z\not=0$ and integrating by parts, we obtain
from (\ref{p2.1}) that
\begin{equation}
\begin{array}{lll}
\displaystyle\int \frac{V^{\prime }(\lambda )\rho _{n}(\lambda )}{\lambda -z}%
d\lambda & = & -\displaystyle\frac{1}{n}\int \frac{\rho _{n}(\lambda )}{%
(\lambda -z)^{2}}d\lambda +\displaystyle\frac{2(n-1)}{n}\int \frac{%
p_{2}^{(n)}(\lambda ,\mu )d\lambda d\mu }{(\lambda -\mu )(\lambda -z)}.%
\end{array}
\label{p2.2}
\end{equation}%
Since $p_{2}^{(n)}(\lambda ,\mu )=p_{2}^{(n)}(\mu ,\lambda )$, we have
\begin{equation*}
2\int \frac{p_{2}^{(n)}(\lambda ,\mu )d\lambda d\mu }{(\lambda -\mu
)(\lambda -z)}=-\int \frac{p_{2}^{(n)}(\lambda ,\mu )d\lambda d\mu }{%
(\lambda -z)(\mu -z)},
\end{equation*}%
and (\ref{p2.2}) takes the form
\begin{eqnarray}  \label{p2.3}
\hspace{-1cm}\int \frac{V^{\prime }(\lambda )\rho _{n}(\lambda )}{\lambda -z}%
d\lambda = & -&\frac{1}{n}\int \frac{\rho _{n}(\lambda )}{(\lambda -z)^{2}}%
d\lambda - \frac{n-1}{n}\int \frac{G_{n}(\lambda ,\mu )}{(\mu -z)(\lambda -z)%
}d\lambda d\mu \\
&-&\frac{n-1}{n}\left( \int \frac{\rho _{n}(\lambda )}{\lambda -z}d\lambda
\right) ^{2},  \notag
\end{eqnarray}%
where $G_{n}(\lambda ,\mu )$ was defined in (\ref{Gnl}). Thus, denoting
\begin{equation}
f_{n}(z)=\int {\frac{\rho _{n}(\lambda )d\lambda }{\lambda -z}}  \label{p2.4}
\end{equation}%
the Stieltjes transform of $\rho_n$, we derive from (\ref{p2.3}) for $%
z=\lambda _{0}+i\eta $, $\eta \not=0$:
\begin{eqnarray}  \label{sq-eq.1}
\frac{n-1}{n}f_{n}^{2}(z)+\int \frac{V^{\prime }(\lambda )\rho _{n}(\lambda )%
}{\lambda -z}d\lambda =-\frac{1}{n}\int \frac{\rho _{n}(\lambda )d\lambda }{%
(\lambda -z)^{2}}-\frac{n-1}{n}\int \frac{G_{n}(\lambda ,\mu )d\lambda d\mu
}{(\mu -z)(\lambda -z)},  \notag
\end{eqnarray}%
and the second integral in the l.h.s. is well defined, since $V$ is linear
for large absolute values of its argument (see the beginning of proof of the
theorem). Moreover, this and (\ref{w_conv}) allow us to pass to the limit $%
n\rightarrow \infty $ in this term. The first term in the r.h.s. of (\ref%
{sq-eq.1}) is $O(n^{-1})$ for any fixed $z$, $\Im z\not=0$. According to (%
\ref{fact}) the second term in the r.h.s. of (\ref{sq-eq.1}) also vanishes
in the limit $n\rightarrow \infty $ and, according to (\ref{w_conv}), $%
f_{n}(z)\rightarrow f(z)$ as $n\rightarrow \infty $ uniformly on a compact
set of $\mathbb{C}\setminus \mathbb{R}$. Therefore, taking the limit $%
n\rightarrow \infty $ in (\ref{sq-eq.1}), we get equation (\ref{sq-eq}).
Setting $z=\lambda +i\eta $, we rewrite the equation as
\begin{equation}
f^{2}(z)+V^{\prime }(\lambda )f(z)+\int \frac{V^{\prime }(\mu )-V^{\prime
}(\lambda )}{\mu -z}N(d\mu) =0.  \label{sq-eq.2}
\end{equation}%
Solving this quadratic equation in $f$ and using the inversion formula for
the Stieltjes transform, we obtain (\ref{sq-rep}) -- (\ref{Q}).

Note that (\ref{sq-rep}) and (\ref{Lip}) imply that $\rho (\lambda )$ is
bounded, because
\begin{equation*}
|Q(\lambda )|\leq \int \frac{|V^{\prime }(\lambda )-V^{\prime }(\mu )|}{%
|\lambda -\mu |}\rho (\mu )d\mu \leq C\int \rho (\mu )d\mu =C.
\end{equation*}%
It is also clear from (\ref{Lip}) and (\ref{sq-rep}) that to prove (\ref%
{d-rho}) it suffices to prove the same inequality for $Q(\lambda )$. To this
end we take any $h>0$ and write
\begin{eqnarray*}
&&|Q(\lambda +h)-Q(\lambda )|\leq \int_{|\lambda -\mu |\leq 2h}\bigg(\frac{%
|V^{\prime }(\lambda )-V^{\prime }(\mu )|}{|\lambda -\mu |}+\frac{|V^{\prime
}(\lambda +h)-V^{\prime }(\mu )|}{|\lambda +h-\mu |}\bigg)\rho (\mu )d\mu \\
&&\hspace{1cm}+\int_{|\lambda -\mu |>2h}\bigg(\frac{|V^{\prime }(\lambda
+h)-V^{\prime }(\lambda )|}{|\lambda +h-\mu |}+\frac{|V^{\prime }(\lambda
)-V^{\prime }(\mu )|h}{|\lambda -\mu |\cdot |\lambda +h-\mu |} \bigg)\rho
(\mu )d\mu \\
&&\hspace{4cm}\leq C\sup_{\lambda \in \mathbb{R}}\rho (\lambda )\,\,h(|\log
h|+1).
\end{eqnarray*}%
Theorem \ref{t:IDS.2} is proved.

\medskip

\textbf{Proof of Theorem \ref{t:IDS.3}.} Set
\begin{equation}
V_{1}(\lambda )=\displaystyle\frac{1}{2}(u(\lambda ;N)-u_{\ast }),\quad
u_{1}(\lambda )=u(\lambda ;N)-V_{1}(\lambda )  \label{V_1}
\end{equation}%
where $u(\lambda ;N)$ and $u_{\ast }$ are defined by (\ref{IDS.1.11}) -- (%
\ref{condin}). It is easy to see that $V_{1}(\lambda )=0$, $\lambda \in
\sigma _{N}$, $V_{1}(\lambda )\geq 0$, $\lambda \not\in \sigma _{N}$, and $%
u_{1}(\lambda )$ attains its minimum $u_{\ast }$ for $\lambda \in \sigma
_{N} $.

Consider the Hamiltonians:
\begin{eqnarray}  \label{IDS.P2.1}
H_{n}^{(1)}(\Lambda)&=&-V_1(\lambda_{1})+ \displaystyle\sum_{i=1}^{n}V(%
\lambda_{i})-\displaystyle\frac{2}{n}\displaystyle\sum_{1\le i<j\le n} \log
|\lambda_{i}-\lambda_{j}|, \\
H_{n}^{(1a)}(\Lambda)&=& \displaystyle\frac{n-1}{n}u_{1}(\lambda_{1})+\frac{1%
}{n}(V(\lambda_1)-V_1(\lambda_1)) +\displaystyle\sum_{i=2}^{n}V(\lambda_{i})
-\displaystyle\frac{2}{n}\displaystyle\sum_{2\leq i<j\le n}\log
|\lambda_{i}-\lambda_{j}| .  \notag
\end{eqnarray}
Denote
\begin{equation*}
p_{n,\beta}^{\natural }(\Lambda)=(Q^{\natural }_{n,\beta})^{-1} \exp
\{-\beta nH_{n}^{\natural }(\Lambda)/2\},\quad \natural=(1),(1a)
\end{equation*}
the corresponding probability densities (cf (\ref{psymb})).

Using the r.h.s inequality in (\ref{Bog}) for $\mathcal{H}_{1}=H_{n}^{(1)}$,
$\mathcal{H}_{2}=H_{n}^{(1a)}$ and $T=2/\beta n$, we get
\begin{equation}  \label{IDS.p2.2}
\ \log Q_{n,\beta}^{(1)}-\log Q_{n,\beta}^{(1a)}\leq I_1+I_2,
\end{equation}
where
\begin{equation*}
\begin{array}{l}
I_1=\beta\displaystyle\sum_{i=2}^n\displaystyle\int \log
|\lambda_{1}-\lambda_{i}|(p _{2}^{(n,1)}(\lambda_{1},\lambda_{i})- \rho
_{n}^{(1)}(\lambda_{1})\rho_{n}^{(2)}(\lambda_{i})) d\lambda_{1}d\lambda_{i},
\\
I_2=\beta(n-1)\displaystyle\int \log |\lambda_{1}-\lambda_{2}| (\overline
N_{n}^{(2)}(d\lambda_{2})-N(d\lambda_2)) \rho
_{n}^{(1)}(\lambda_{1})d\lambda_{1},%
\end{array}%
\end{equation*}
$\rho _{n}^{(1)}$, and $\rho _{n}^{(2)}$ are the first marginal densities
corresponding to $\lambda_{1}$ and $\lambda_{i}$, $i=2,\dots,n$ for the
Hamiltonian $H_{n}^{(1)}$, $\overline N_{n}^{(\alpha)}(d\lambda)=\rho
_{n}^{(\alpha)}(\lambda)d\lambda$, $\alpha=1,2$ (note that $\rho
_{n}^{(1)}\not=\rho _{n}^{(2)}$ since $H_{n}^{(1)}$ is not symmetric in $%
\lambda_{1}$ and $\lambda_{i},\,i=2,\dots, n$), $p _{2}^{(n,1)}$ and $p
_{2}^{(n,2)} $ are the second marginal densities, corresponding to $%
\lambda_{1},\lambda_{i},i=2,\dots,n$ and $\lambda_{i},\lambda_{j},2\le
i<j\le n$ (note that $p _{2}^{(n,1)}\not=p _{2}^{(n,2)}$ and $p _{2}^{(n,1)}$
is not symmetric because of the same reason).

Repeating the argument that leads to formulas (\ref{b_above}) and (\ref%
{b_below}) below, we get analogs of (\ref{exp_est}) for $\rho _{n}^{(1)}$
and $\rho _{n}^{(1)}$ that allow us to restrict integration in the r.h.s. of
(\ref{IDS.p2.2}) to $[-1/2,1/2]$. Besides, using (\ref{Four_l}) for $\log
|\lambda -\mu |^{-1}$ we obtain similarly to (\ref{IDS.t1.2}) and (\ref%
{IDS.t1.3})
\begin{eqnarray}
&&\hspace{-1cm}|I_{1}|\leq O(e^{-nc})+\beta \bigg|\displaystyle%
\sum_{|k|<M}l^{(k)}\bigg\langle e^{ikr\lambda _{1}}\displaystyle%
\sum_{j=2}^{n}(e^{ik\pi \lambda _{j}}-\langle e^{ikr\lambda _{j}}\rangle )%
\bigg\rangle\bigg|+|R_{M}|  \label{IDS.p2.3} \\
&&\leq \beta \bigg[\displaystyle\sum_{|k|<M}l^{(k)}\bigg]^{1/2}\bigg[%
\displaystyle\sum_{|k|<M}l^{(k)}\bigg\langle\bigg|\displaystyle%
\sum_{j=2}^{n}(e^{ik\pi \lambda _{j}}-\langle e^{ik\pi \lambda _{j}}\rangle )%
\bigg|^{2}\bigg\rangle\bigg]^{1/2}+|R_{M}|  \notag \\
&&\leq C\log ^{1/2}M\bigg[O(\log ^{1/2}M)+n^{2}\displaystyle%
\int_{-1}^{1}\log \frac{1}{|\lambda -\mu |}G_{n}^{(2)}(\lambda ,\mu
)d\lambda d\mu -R_{M}^{(1)}\bigg]^{1/2},  \notag
\end{eqnarray}%
where we denote (cf (\ref{Gnl})
\begin{equation*}
\langle \dots \rangle :=\int (\dots )p_{n,\beta }^{(1)}(\Lambda )d\Lambda
,\quad G_{n}^{(2)}(\lambda ,\mu )=p_{2}^{(n,2)}(\lambda ,\mu )-\rho
_{n}^{(2)}(\lambda )\rho _{n}^{(2)}(\mu ),
\end{equation*}%
$M=n^{2+6/\gamma }$ and $R_{M}$ and $R_{M}^{(1)}$ are the remainder terms
which are the contributions of sums from $|j|=M+1$ to $\infty $ in the
Fourier series (see (\ref{IDS.t1.3}) for the estimate of such terms).

Likewise, considering $H_{a}^{(1)}$ of the form (\ref{IDS.2.8}) with $%
V(\lambda _{1})$ replaced by $V(\lambda _{1})-V_{1}(\lambda _{1})$ and
repeating the arguments, leading to (\ref{b_G}) and (\ref{b_L}), we obtain
analogs of these inequalities for the Hamiltonian $H_{n}^{(1)}$:
\begin{equation}
\begin{array}{c}
\displaystyle\int \log \displaystyle\frac{1}{|\lambda -\mu |}%
G_{n}^{(2)}(\lambda ,\mu )d\lambda d\mu =O(\displaystyle\frac{\log n}{n}),
\\
0\leq \mathcal{L}[\overline{N}_{n}^{(2)}-N,\overline{N}_{n}^{(2)}-N]\leq %
\displaystyle\frac{C\log n}{n}.%
\end{array}
\label{IDS.p2.4}
\end{equation}%
This and (\ref{IDS.p2.3}) yield $I_{1}=O(n^{1/2}\log n)$. Similarly, on the
basis of the second line of (\ref{IDS.p2.4}) and the Schwarz inequality we
get $I_{2}=O(n^{1/2}\log n)$. Plugging these estimates in (\ref{IDS.p2.2}),
we obtain
\begin{equation}
\log Q_{n,\beta }^{(1)}-\log Q_{n,\beta }^{(1a)}\leq Cn^{1/2}\log n.
\label{IDS.p2.7}
\end{equation}%
Now we use the r.h.s inequality in (\ref{Bog}) for $\mathcal{H}%
_{1}=H_{n}^{(1a)}$, $\mathcal{H}_{2}=H_{n}$ and $T=2/\beta n $ to get the
bound
\begin{eqnarray}
\log Q_{n,\beta }^{(1a)}&-&\log Q_{n,\beta }\leq \beta n \displaystyle\int
V_{1}(\lambda )\rho _{n}^{(1a)}(\lambda )d\lambda  \label{IDS.p2.8} \\
&+&\beta (n-1) \displaystyle\int \log |\lambda _{1}-\lambda _{2}|(\rho
_{n}^{(2a)}(\lambda _{2})d\lambda _{2}-N(d\lambda _{2}))\rho
_{n}^{(1a)}(\lambda _{1})d\lambda _{1},  \notag
\end{eqnarray}%
where $\rho _{n}^{(1a)}$ and $\rho _{n}^{(2a)}$ are the first marginal
densities of the Hamiltonian $H_{n}^{(1a)}$, corresponding to $\lambda _{1}$
and $\lambda _{i},i=2,\dots ,n$. It is easy to see that (cf (\ref{rho_a}))
\begin{equation}
\rho _{n}^{(1a)}(\lambda )={\frac{\exp \{\beta \lbrack -(n-1)u_{1}(\lambda
)/2+V_{1}(\lambda )-V(\lambda )]\}}{\int \exp \{\beta \lbrack
-(n-1)u_{1}(\lambda )/2+V_{1}(\lambda )-V(\lambda )]\}d\lambda }}.
\label{rho^1a}
\end{equation}%
According to definitions (\ref{V_1}) and (\ref{IDS.1.11}) $V_{1}(\lambda )=0$
for $\lambda \in \sigma _{N}$ and in view of (\ref{cond0}) and Proposition %
\ref{p:enpr}) (see (\ref{Holdu})), the function $V_{1}-V$ admits the bounds:
\begin{equation*}
\begin{array}{l}
V_{1}(\lambda )-V(\lambda )\leq C,\quad \lambda \in \mathbb{R}, \\
V_{1}(\lambda )-V(\lambda )\geq -C,\quad \lambda \in \sigma _{N}.%
\end{array}%
\end{equation*}%
Besides, the integral in the denominator of (\ref{rho^1a}) is bounded from
below by the integral over $\sigma _{N}$, which is bounded from below by $%
|\sigma _{N}|\exp \{-\beta (n-1)u_{\ast }/2-C\}$ and according to Theorem %
\ref{t:IDS.2} $|\sigma _{N}|\not=0$, where $|\sigma _{N}|$ is the Lebesgue
measure of $\sigma _{N}$. Taking into account the above bounds, and denoting
$I_{1}^{\prime }$ the first term in the r.h.s. of (\ref{IDS.p2.8}), we
obtain
\begin{equation*}
|I_{1}^{\prime }|\leq e^{2C}d_{n}/|\sigma _{N}|,
\end{equation*}%
where $d_{n}$ is defined in (\ref{e_n}).

The second term in the r.h.s. of (\ref{IDS.p2.8}) can be estimated by
Schwarz inequality (\ref{LSchw}):
\begin{equation*}
\begin{array}{l}
\displaystyle\int \log |\lambda _{1}-\lambda _{2}|(\rho _{n}^{(2a)}(\lambda
_{2})d\lambda _{2}-N(d\lambda _{2}))\rho _{n}^{(1a)}(\lambda _{1})d\lambda
_{1} \\
\quad =-\mathcal{L}[\rho _{n}^{(2a)}d\lambda -N,\rho _{n}^{(1a)}d\lambda
]\leq \mathcal{L}^{1/2}[\rho _{n}^{(2a)}d\lambda -N,\rho _{n}^{(2a)}d\lambda
-N]\mathcal{L}^{1/2}[\rho _{n}^{(1a)}d\lambda ,\rho _{n}^{(1a)}d\lambda ].%
\end{array}%
\end{equation*}%
According to the above $\rho _{n}^{(1a)}$ is bounded and decays at infinity
as $C_{1}\exp \{-nC_{2}V(\lambda )\}$, hence the second factor is bounded.
To estimate the first factor we note that $\rho _{n}^{(2a)}$ coincides with
the first marginal density of the Hamiltonian
\begin{equation*}
H_{n}^{^{\prime }}(\lambda _{2},...,\lambda _{n})=\sum_{i=2}^{n}V(\lambda
_{i})-\frac{2}{n}\sum_{2\leq i<j}\log |\lambda _{i}-\lambda _{j}|.
\end{equation*}%
Thus, the bound for the second factor follows from (\ref{w_conv}) with $\rho
_{n}$ replaced by $\rho _{n}^{(2a)}$. Finally, from (\ref{IDS.p2.7}) and (%
\ref{IDS.p2.7}) we derive
\begin{equation}
\log \frac{Q_{n,\beta }^{(1)}}{Q_{n,\beta }}=\log \frac{Q_{n,\beta }^{(1)}}{%
Q_{n,\beta }^{(1a)}}+\log \frac{Q_{n,\beta }^{(1a)}}{Q_{n,\beta }}\leq
\,C\,(n^{1/2}\log n+nd_{n}).  \label{IDS.p2.9}
\end{equation}%
The assertion of Theorem \ref{t:IDS.3} follows.


\subsection{Auxiliary results}

\textbf{Proof of Lemma~\ref{l:Bog}.} Define $F:[0,1]\rightarrow \mathbb{R}%
_{+}$ as
\begin{equation*}
F(t)=T\log \int \exp \{-T^{-1}((1-t)\mathcal{H}_{1}(\Lambda )-t\mathcal{H}%
_{2}(\Lambda ))\}d\Lambda
\end{equation*}%
It is evident that $F^{\prime \prime }(t)\geq 0$. Therefore we have for all $%
t\in \lbrack 0,1]$:
\begin{equation*}
F^{\prime }(0)\leq F^{\prime }(t)\leq F^{\prime }(1),
\end{equation*}%
and integrating with respect to $t$, we get
\begin{equation*}
F^{\prime }(0)\leq F(1)-F(0)\leq F^{\prime }(1).
\end{equation*}%
Inequality (\ref{Bog}) follows.

\medskip

\textbf{Proof of Theorem \ref{t:IDS.t1} (i).} We prove first that there
exists some $n$-independent $C$, such that
\begin{equation}
\int \rho _{n}(\lambda )V(\lambda )d\lambda \leq C.  \label{rho,V}
\end{equation}%
Choosing in (\ref{Bog}) $T=2/\beta n$, $H_{1}=H$ and $H_{1}=H^{(\epsilon )}$%
, where $H^{(\epsilon )}$ has the form (\ref{IDS.1.7}) with a function $V$
replaced by $(1-\epsilon _{1})V$, $\epsilon _{1}=\epsilon /2(1+\epsilon )$,
we get from (\ref{Bog})
\begin{equation*}
\epsilon _{1}\int \rho _{n}(\lambda )V(\lambda )d\lambda \leq \frac{2}{%
n^{2}\beta }\log Q_{n,\beta }^{(\epsilon )}-\frac{2}{n^{2}\beta }\log
Q_{n,\beta }.
\end{equation*}%
Now (\ref{rho,V}) follows from the inequalities:
\begin{equation*}
\displaystyle\frac{2}{n^{2}\beta }\log Q_{n,\beta }^{(\epsilon )}\leq
-m,\quad \displaystyle\frac{2}{n^{2}\beta }\log Q_{n,\beta }\geq -M-3/2
\end{equation*}%
with $M$ and $m$ defined in (\ref{M}). The first can be easily obtained by
the Laplace method, if we use the bound
\begin{equation}
\log |\lambda -\mu |\leq \log (1+|\lambda |)+\log (1+|\mu |),  \label{b_log}
\end{equation}%
and the fact that $(1-\epsilon _{1})V$ satisfies (\ref{cond0}). And the
second follows from the Jensen inequality:
\begin{equation*}
Q_{n,\beta }^{(\epsilon )}>\int_{-1/2}^{1/2}\dots \int_{-1/2}^{1/2}d\Lambda
\exp \{-H(\Lambda )\}\geq \exp \bigg\{-\int_{-1/2}^{1/2}\dots
\int_{-1/2}^{1/2}H(\Lambda )d\Lambda \bigg\}\geq e^{\beta n^{2}C_{1}/2}
\end{equation*}%
with $\Lambda =(\lambda _{1},\dots ,\lambda _{n})$ and%
\begin{equation*}
C_{1}=-\int_{-1/2}^{1/2}|V(\lambda )|d\lambda +\int_{-1/2}^{1/2}\log
|\lambda -\mu |d\lambda d\mu \geq -M-3/2.
\end{equation*}%
Denote for the moment $H(\lambda _{1},\dots ,\lambda _{n})$ of (\ref{IDS.1.7}%
) as $H_{n}(\lambda _{1},\dots ,\lambda _{n})$. Then (\ref{IDS.1.7}) implies
\begin{eqnarray}
\frac{\beta n}{2}H_{n}(\lambda _{1},\dots ,\lambda _{n}) &=&\frac{\beta (n-1)%
}{2}H_{n-1}(\lambda _{2},\dots ,\lambda _{n})+\frac{\beta n}{2}V(\lambda
_{1})  \label{HnHn-1} \\
&+&\frac{\beta }{2}\sum_{i=2}^{n}(V(\lambda _{i})-2\log |\lambda
_{1}-\lambda _{i}|),  \notag
\end{eqnarray}%
and in view of (\ref{b_log}) and (\ref{cond0}) \ we obtain%
\begin{equation*}
\frac{\beta n}{2}H_{n}(\lambda _{1},\dots ,\lambda _{n})\geq \frac{\beta
(n-1)}{2}H_{n-1}(\lambda _{2},\dots ,\lambda _{n})+\frac{\beta n}{2}%
V(\lambda _{1})-\beta (n-1)\log (1+|\lambda _{1}|).
\end{equation*}%
This and (\ref{Qnb}) yield:%
\begin{eqnarray}
&&\hspace{-2cm}\int Q_{n-1,\beta }^{-1}\exp \Big\{-\frac{\beta (n-1)}{2}%
H_{n-1}(\lambda _{2},...,\lambda _{n})\Big\}d\lambda _{2}\dots d\lambda _{n}
\label{b_above} \\
&\leq &\exp \Big\{-\frac{\beta n}{2}V(\lambda _{1})+\beta (n-1)\log
(1+|\lambda _{1}|)\Big\}  \notag \\
&\leq &\exp \Big\{-\frac{\beta n\epsilon }{2(1+\epsilon )}V(\lambda _{1})%
\Big\}.  \notag
\end{eqnarray}%
On the other hand, by using again (\ref{HnHn-1}) and the Jensen inequality
for the "Gibbs" measure $e^{-\beta (n-1)H_{n-1}/2}Q_{n-1,\beta }^{-1}$, we
obtain
\begin{equation*}
Q_{n-1,\beta }^{-1}Q_{n,\beta }\geq \int_{-1/2}^{1/2}d\lambda _{1}\exp %
\bigg\{-\frac{\beta n}{2}u_{n}(\lambda _{1};\overline{N}_{n-1})-\frac{\beta
(n-1)}{2}\int V(\lambda )\rho _{n-1}(\lambda )d\lambda \bigg\},
\end{equation*}%
where $\overline{N}_{n-1}$ is defined in (\ref{Nbp}) -- (\ref{Nnrho}) and $%
u_{n}$ is defined in (\ref{u_n}).

Using the Jensen inequality with respect to $\nu _{0}$, the Lebesgue measure
on the interval $[-1/2,1/2]$, we get further
\begin{equation}\label{b_below}
\hspace{-1cm}Q_{n-1,\beta }^{-1}Q_{n,\beta }\geq e^{-(n-1)\beta C/2}\exp %
\Big\{-\frac{n\beta }{2}\int_{-1/2}^{1/2}V(\lambda )d\lambda -(n-1)\beta
\mathcal{L}[\nu _{0},\overline{N}_{n-1}]\Big\}.
\end{equation}%
where $C$ is defined in (\ref{rho,V}). But since
\begin{equation}
-\mathcal{L}(\lambda ;\nu _{0})=(1/2-\lambda )\log (1/2-\lambda
)+(1/2+\lambda )\log (1/2+\lambda )-1\geq -1-\log 2,  \label{b_log.1}
\end{equation}%
we have
\begin{equation*}
-\mathcal{L}[\nu _{0},\overline{N}_{n-1}]=-\int \mathcal{L}(\lambda ;\nu
_{0})\overline{N}_{n-1}(d\lambda )\geq -1-\log 2,
\end{equation*}%
hence
\begin{equation}
Q_{n-1,\beta }^{-1}Q_{n,\beta }\geq e^{-n\beta C_{1}/2},\quad
C_{1}=C+2+2\log 2+M,  \label{ti-C}
\end{equation}%
and%
\begin{equation}
\begin{array}{rcl}
\rho _{n}(\lambda ) & = & \dfrac{Q_{n-1,\beta }}{Q_{n,\beta }}Q_{n-1,\beta
}^{-1}\displaystyle\int e^{-\beta nH(\lambda _{1},\dots \lambda _{n})/2}%
\label{L}d\lambda _{2}\dots d\lambda _{n} \\
& \leq  & e^{\beta nC_{1}}e^{-\beta n\epsilon V(\lambda )/2(1+\epsilon
)}\leq e^{-\beta n\epsilon V(\lambda )/4(1+\epsilon )},\quad |\lambda |>L,%
\end{array}%
\end{equation}%
if $L$ is big enough. This proves the first bound in (\ref{exp_est}). The
bound (\ref{exp_est}) for $p_{2,\beta }^{(n)}$ can be proved analogously.

\medskip

\textbf{Proof of Lemma~\ref{l:IDS.l2}.} (i) Using (\ref{Lpos}), it is easy
to see that $\Phi _{n}(m)$ is convex, ~i.e.,
\begin{equation}
\Phi _{n}\bigg[\frac{m^{(1)}+m^{(2)}}{2}\bigg]\leq \frac{\Phi
_{n}[m^{(1)}]+\Phi _{n}[m^{(2)}]}{2}.  \label{conv}
\end{equation}%
Let us show that $\Phi _{n}[m]$ is bounded from below. Let $\nu _{0}$ be the
Lebesgue measure on the interval $[-1/2,1/2]$. Then, using the Jensen
inequality and then (\ref{b_log.1}), we get similarly to (\ref{b_below}) -- (%
\ref{ti-C})%
\begin{eqnarray}
&&\hspace{-1.5cm}\frac{2}{\beta n}\log \int_{-1/2}^{1/2}\exp \bigg\{-\frac{%
\beta n}{2}V(\lambda )-\beta (n-1)\mathcal{L}(\lambda ,m)\bigg\}d\lambda
\label{b_b} \\
&\geq &-\int \mathcal{L}(\lambda ;\nu _{0})m(d\lambda
)-\int_{-1/2}^{1/2}V(\lambda )d\lambda >-(1+\log
2)-\int_{-1/2}^{1/2}V(\lambda )d\lambda .  \notag
\end{eqnarray}%
Combining this inequality with (\ref{Lpos}) we conclude that $\inf \Phi
_{n}[m]>-\infty $.

Consider a minimizing sequence $\{m^{(k)}\}$ of measures, satisfying (\ref%
{cond_m}) and such that
\begin{equation*}
\lim_{k\rightarrow \infty }\Phi _{n}[m^{(k)}]=\inf_{m\in \mathcal{C}^{\ast
}}\Phi _{n}[m]:=\Phi _{n}^{\ast }.
\end{equation*}%
Then for any $\varepsilon >0$ there exists $k_{\varepsilon }$ such that
\begin{equation*}
\Phi _{n}^{\ast }+\varepsilon \geq \Phi _{n}[m^{(k)}]\geq \Phi _{n}^{\ast
},\quad k>k_{\varepsilon }.
\end{equation*}%
This and (\ref{conv}) yield for $k,l>k_{\varepsilon }$,
\begin{equation*}
\Phi _{n}^{\ast }+\varepsilon \geq \displaystyle\frac{\Phi
_{n}[m^{(k)}]+\Phi _{n}[m^{(l)})]}{2}\geq \Phi _{n}\bigg[\frac{%
m^{(k)}+m^{(l)}}{2}\bigg]\geq \Phi _{n}^{\ast }.
\end{equation*}%
Besides, it follows from (\ref{IDS.2.9}) that
\begin{equation*}
\Phi _{n}[m]=\frac{n-1}{n}\mathcal{L}[m,m]+\Psi _{n}[m],
\end{equation*}%
where $\Psi _{n}$ is also convex. Using the convexity of $\Psi _{n}$ and the
previous inequality, we obtain:%
\begin{equation}
\mathcal{L}[m^{(k)}-m^{(l)},m^{(k)}-m^{(l)}]\leq 4\bigg(\frac{\Phi
_{n}[m^{(k)}]+\Phi _{n}[m^{(l)}]}{2}-\Phi _{n}\bigg[\frac{m^{(k)}+m^{(l)}}{2}%
\bigg]\bigg)\leq 2\varepsilon .  \label{IDS.3.14}
\end{equation}%
In other words, the sequence $\{m^{(k)}\},\;m^{(k)}\subset \mathcal{C}^{\ast
}$ of (\ref{cond_m}) satisfies the Cauchy condition with respect to the norm
$||m||_{\ast }=\mathcal{L}^{1/2}[m,m]$ and, as a result, has a limit point $%
m_{n}$ in this cone by Proposition \ref{p:enpr} (v). This point $m_{n}$ is a
minimum point for $\Phi _{n}$. Besides, since the second derivative of $\Phi
_{n}$ in any direction is bounded from below by a positive constant, $m_{n}$
is a unique minimum point.

Consider the measure $N_{n}^{(a)}(d\lambda ;m_{n})$, defined by (\ref{Nan})
for $m=m_n$. Taking the derivative of $\Phi _{n}(m_{n}+t(m-m_{n}))$ with
respect to $t$ at $t=0$ it is easy to find that for any $m\in \mathcal{C}%
^{\ast }$
\begin{equation}
\mathcal{L}[m_{n}-N_{n}^{(a)},m-m_{n}]=0.  \label{der_Phi}
\end{equation}%
Let us show that for any $m\in \mathcal{C}^{\ast }$ we have for $\rho
_{n}^{(a)}$ of (\ref{rho_a})
\begin{equation}
\rho _{n}^{(a)}(\lambda ;m)\leq e^{-nCV(\lambda )},\quad |\lambda |>1/2.
\label{exp_est2}
\end{equation}%
Since $\hbox{supp}\,\,m\subset \lbrack -1/2,1/2]$ and $m(\mathbb{R})\leq 1$,
we have 
\begin{equation*}
-\mathcal{L}(\lambda ;m)\leq \log (1+|\lambda |).
\end{equation*}%
This and (\ref{cond0}) yield for the numerator of (\ref{rho_a})
\begin{equation*}
e^{-\beta nu_n(\lambda,m )}\leq e^{-\beta n \epsilon V(\lambda
)/2(1+\epsilon )}.
\end{equation*}%
To estimate the denominator in (\ref{rho_a}) we use (\ref{b_b}). Then, using
the last inequality in (\ref{L}) we get (\ref{exp_est2}). We recall here
that we use the scaling of the $\lambda $-axis such that $L<1/2$.

Consider now the measure
\begin{equation*}
\widetilde{N}_{n}^{(a)}(\lambda )=N_{n}^{(a)}(\lambda ;m_{n})\mathbf{1}%
_{|\lambda |<1/2}.
\end{equation*}%
It follows from (\ref{exp_est2}) that
\begin{equation}
|\mathcal{L}(\lambda ;\widetilde{N}_{n}^{(a)})-\mathcal{L}(\lambda
;N_{n}^{(a)})|\leq e^{-nc}/2.  \label{exp_est3}
\end{equation}%
Thus
\begin{equation*}
\mathcal{L}[N_{n}^{(a)}-\widetilde{N}_{n}^{(a)},N_{n}^{(a)}-\widetilde{N}%
_{n}^{(a)}]\leq e^{-nc}/2.
\end{equation*}%
Besides, replacing in (\ref{der_Phi}) $m$ by $\widetilde{N}_{n}^{(a)}$, we
get
\begin{equation*}
\mathcal{L}[m_{n}-N_{n}^{(a)},m_{n}-\widetilde{N}_{n}^{(a)}]=0,
\end{equation*}%
hence (\ref{exp_est4}) follows.

(ii) Define (cf (\ref{IDS.2.9}))
\begin{equation}\label{Phi^1}
\Phi _{n}^{(1)}[m]=\frac{(n-1)}{n}\mathcal{L}[m,m]+\frac{2}{\beta n}\log
\int_{-1/2}^{1/2}d\lambda e^{-\beta nu_{n}(\lambda ;m)/2},
\end{equation}%
where $u_{n}$ is given by (\ref{u_n}). Then (\ref{exp_est2}) implies for any
$\Phi _{n}$ of (\ref{IDS.2.9}) and $m\in \mathcal{C}^{\ast }$ of (\ref%
{cond_m})%
\begin{equation}
\left\vert \Phi _{n}^{(1)}[m]-\Phi _{n}[m]\right\vert \leq e^{-nc}.
\label{d_Phi}
\end{equation}%
Repeating the proof of existence of a minimizer $\Phi _{n}$ in (i), we
obtain that there exists a unique measure $m_{n}^{(1)}\in \mathcal{C}^{\ast
} $ such that
\begin{equation}
\Phi _{n}^{(1)}[m_{n}^{(1)}]=\inf_{m\in \mathcal{C}^{\ast }}\Phi
_{n}^{(1)}[m].  \label{m^1}
\end{equation}%
Now, if we define (cf (\ref{rho_a}), (\ref{Nan}))
\begin{equation}
N_{n}^{(a,1)}(d\lambda )=e^{-\beta nu_{n}(\lambda ;m_{n}^{(1)})/2}\mathbf{1}%
_{|\lambda |\leq 1/2}\Big(\int_{-1/2}^{1/2}e^{-\beta nu_{n}(\lambda
;m_{n}^{(1)})/2} d\lambda\Big)^{-1},
\end{equation}%
then the analog of (\ref{der_Phi}) for $\Phi _{n}^{(1)}$ implies the
equation
\begin{equation}
m_{n}^{(1)}=N_{n}^{(a,1)}.  \label{sce}
\end{equation}%
Consider $F:[0,1]\rightarrow \mathbb{R}$, given by
\begin{eqnarray}
F(t) &=&\frac{(n-1)}{n}\mathcal{L}%
[m_{n}^{(1)}+t(N-m_{n}^{(1)}),m_{n}^{(1)}+t(N-m_{n}^{(1)})]  \label{F} \\
&&-(1-t)\int u_{n}(\lambda ;m_{n}^{(1)})m_{n}^{(1)}(d\lambda )-t\int
u(\lambda ;N)N(d\lambda ),  \notag
\end{eqnarray}%
where $u$ and $u_{n}$ are defined in (\ref{IDS.1.11}) and (\ref{u_n}). It is
evident, that $F^{\prime \prime }(t)\geq 0$ and we obtain in view of (\ref%
{IDS.1.10}) -- (\ref{IDS.1.11})%
\begin{eqnarray}
F(1)-F(0) &\leq &F^{\prime }(1)=2\frac{n-1}{n}\mathcal{L}[N,N-m_{n}^{(1)}]
\label{n1} \\
&+&\int u_{n}(\lambda ;m_{n}^{(1)})m_{n}^{(1)}(d\lambda )-\int V(\lambda
)N(d\lambda )-2\mathcal{L}[N,N]  \notag \\
&=&\int u_{n}(\lambda ;m_{n}^{(1)})m_{n}^{(1)}(d\lambda )-\int u_{n}(\lambda
;m_{n}^{(1)})N(d\lambda )+O(n^{-1}).  \notag
\end{eqnarray}%
This inequality and (\ref{d_Phi}) imply
\begin{eqnarray}
0 &\leq &\Phi _{n}[N]-\Phi _{n}[m_{n}]=(\Phi _{n}^{(1)}[N]-F(1))+(F(1)-F(0))+
\label{n2} \\
&+&(F(0)-\Phi _{n}^{(1)}[m_{n}^{(1)}])+O(e^{-nc})\leq (\Phi
_{n}^{(1)}[N]-F(1))  \notag \\
&+&\int u_{n}(\lambda ;m_{n}^{(1)})m_{n}^{(1)}(d\lambda )-\int u_{n}(\lambda
;m_{n}^{(1)})N(d\lambda )+(F(0)-\Phi _{n}^{(1)}[m_{n}^{(1)}])+O(n^{-1}),
\notag
\end{eqnarray}%
where $\Phi _{n}^{(1)}(m)$ and $m_{n}^{(1)}$ are defined in (\ref{Phi^1})
and (\ref{m^1}).

Therefore to prove (\ref{est}) it suffices to have the inequalities:
\begin{equation}
\begin{array}{l}
\Phi _{n}^{(1)}[N]-F(1)\leq 0; \\
\displaystyle\int u_{n}(\lambda ;m_n^{(1)})m_n^{(1)}(d\lambda )-\displaystyle%
\int u_{n}(\lambda ;m_n^{(1)})N(d\lambda )\leq Cn^{-1}\log n; \\
F(0)-\Phi _{n}^{(1)}[m_n^{(1)}]\leq Cn^{-1}\log n.%
\end{array}
\label{n3}
\end{equation}%
The first inequality follows from (\ref{exp_est2}), (\ref{condin}) -- (\ref%
{u_*}) and the simple bound
\begin{equation*}
\frac{2}{\beta n}\log \int_{-1/2}^{1/2}e^{-\beta n u(\lambda ;N)/2}d\lambda
\leq -\min_{\lambda \in \mathbb{R}}u(\lambda ;N)=-\int u(\lambda
;N)N(d\lambda ),
\end{equation*}%
where the last equality follows from (\ref{condin}).

To prove the second inequality in (\ref{n3}) we introduce the function
\begin{equation}  \label{dnl1}
\delta _{n}(\lambda )=\frac{n^{1/\gamma}}{2}\mathbf{1}_{|\lambda
|<n^{-1/\gamma}},
\end{equation}%
and consider the convolution operator $\delta^{\ast}_n$ defined for any
finite measure $m$ as
\begin{equation*}
(\delta_{n}^\ast m)(\Delta )=\int_{\Delta }\frac{n^{1/\gamma}}{2}(m(\lambda
+n^{-1/\gamma})-m(\lambda -n^{-1/\gamma}))d\lambda ,\quad m(\lambda
)=m((-\infty ,\lambda ]).
\end{equation*}%
It is evident that for any non-negative measure $m$ such that $m(\mathbb{R}%
)\leq 1$ the measure $\delta_{n}^\ast m$ has a density bounded by $%
n^{1/\gamma}$. This implies, in particular, that
\begin{equation}  \label{n4}
\bigg|\mathcal{L}(\lambda +h;\delta^\ast_{n}m)-\mathcal{L}(\lambda
;\delta^\ast_{n}m)\bigg| \le n^{1/\gamma}\int_{-1}^{1}\bigg|\log \bigg|1+%
\frac{h}{\lambda -\mu }\bigg|\bigg|d\lambda \leq Ch^{1/2}n^{1/\gamma}.
\end{equation}%
Besides, if the measure $m$ is absolutely continuous and its density is $%
\rho $, then $\delta^\ast_{n}m$ has the density
\begin{equation*}
(\delta _{n}\ast \rho )(\lambda )=\int \delta _{n}(\lambda -\mu )\rho (\mu
)d\mu ,
\end{equation*}%
the convolution of $\delta _{n}$ and $\rho $. We will also use below the
following estimate valid for any function $v:\mathbb{R}\rightarrow \mathbb{C}
$, satisfying the H\"{o}lder condition with the exponent $\gamma $:
\begin{equation}
|(\delta _{n}\ast v)(\lambda )-v(\lambda )|\leq
(2n^{-1/\gamma})^{-1}\int_{|\mu |\leq n^{-1/\gamma}}|v(\lambda +\mu
)-v(\lambda )|d\mu \leq Cn^{-1/\gamma}.  \label{n4a}
\end{equation}%
Moreover, for any $m$ with finite energy (\ref{IDS.1.10}) we have
\begin{equation}
\int \delta _{n}(\lambda -\mu )\mathcal{L}(\mu ;m)d\mu =\mathcal{L}(\lambda
;\delta^\ast_{n}m),  \label{n5}
\end{equation}%
and in view of the relations
\begin{equation*}
\widehat{\delta }_{n}(p):=\int e^{ip\lambda }\delta _{n}(\lambda )d\lambda =%
\frac{\sin pn^{-1/\gamma}}{pn^{-1/\gamma}},\quad |\widehat{\delta }%
_{n}(p)|\leq 1,
\end{equation*}%
and (\ref{reprL}) we obtain
\begin{equation*}
\mathcal{L}[\delta^\ast_{n}m-m,m]=\int_{0}^{\infty }\frac{|\widehat{m}%
(p)|^{2}}{p}(\widehat{\delta }_{n}(p)-1)dp\leq 0,
\end{equation*}%
hence
\begin{equation}
\mathcal{L}[\delta^\ast_{n}m,m]\leq \mathcal{L}[m,m].  \label{n6}
\end{equation}%
Now we are ready to prove the second and the third inequality in (\ref{n3}).
Using (\ref{sce}), (\ref{Bog}) with $\mathcal{H}_{1}=u_{n}(\lambda;m_n^{(1)}
)$, \; $\mathcal{H}_{2}=0$ and $T=2/(\beta n)$, and then (\ref{d_Phi}) we
have
\begin{eqnarray}
&&\hspace{-1.5cm}-\int u_{n}(\lambda ;m_n^{(1)})m_n^{(1)}(d\lambda )=-\int
u_{n}(\lambda ;m_n^{(1)})N_{n}^{(a,1)}(d\lambda )  \label{n7a} \\
&\geq &\frac{2}{\beta n }\log \int_{-1/2}^{1/2}e^{-\beta n u_{n}(\lambda
;m_n^{(1)})/2}d\lambda -\frac{2}{\beta n }\log \int_{-1/2}^{1/2}d\lambda
\notag \\
&=&\frac{2}{\beta n }\log \int e^{-\beta n u_{n}(\lambda
;m_n^{(1)})/2}d\lambda+O(e^{-nc}).  \notag
\end{eqnarray}%
Besides, we have by Jensen inequality
\begin{eqnarray}
\frac{2}{\beta n }\log \int e^{-\beta n u_{n}(\lambda ;m_n^{(1)})/2}d\lambda
&=&\frac{2}{\beta n }\log \int \delta _{n}(\lambda -\mu )e^{-\beta n
u_{n}(\lambda ;m_n^{(1)})/2}d\mu d\lambda  \label{n7b} \\
&\geq &\frac{2}{\beta n }\log \int e^{-\beta n\check{u}_{n}(\mu )/2} d\mu,
\notag
\end{eqnarray}%
where
\begin{equation*}
\check{u}_{n}(\lambda )=(\delta _{n}\ast V)(\lambda )+ 2\frac{n-1}{n}%
\mathcal{L}(\lambda ;\delta^\ast_{n}m_n^{(1)}).
\end{equation*}%
Observe also that if
\begin{equation*}
\check{u}_{n}^{\ast }:=\min_{\lambda \in \mathbb{R}}\check{u}_{n}(\lambda )=%
\check{u}_{n}(\lambda ^{\ast }),
\end{equation*}%
then (\ref{n4}) with $h=n^{-6/\gamma}$ implies
\begin{equation*}
\check{u}_{n}(\lambda )<\check{u}_{n}^{\ast }+Cn^{-2},\quad |\lambda
-\lambda ^{\ast }|\leq n^{-6/\gamma},
\end{equation*}%
thus
\begin{equation*}
\int d\lambda e^{-\beta n \check{u}_{n}(\lambda )/2}\geq
n^{-6/\gamma}e^{-\beta n \check{u}_{n}^{\ast }/2-Cn^{-1}}.
\end{equation*}%
This bound, (\ref{n7a}), and (\ref{n7b}) yield
\begin{eqnarray}
&&\hspace{-1.5cm}-\int u_{n}(\lambda ;m_n^{(1)})m_n^{(1)}(d\lambda )
\label{n7} \\
&\geq &\frac{2}{\beta n}\log \int_{-1/2}^{1/2}e^{-\beta n \check{u}%
_{n}(\lambda )/2}d\lambda +O(e^{-nc})\geq -\check{u}_{n}^{\ast }-Cn^{-1}\log
n.  \notag
\end{eqnarray}%
Using this inequality and (\ref{n4a}) for $v(\lambda )=\mathcal{L}(\lambda
;N)$ and $v(\lambda )=V(\lambda )$, we obtain in view of (\ref{Holdu}), (\ref%
{u_n}), and (\ref{dnl1})%
\begin{eqnarray*}
&&\hspace{-0.5cm}-\int u_{n}(\lambda; m_n^{(1)})m_n^{(1)}(d\lambda ) \geq
-\int \check{u}_{n}(\lambda )N(d\lambda )-Cn^{-1}\log n \\
&&\hspace{0.5cm}= -2\frac{n-1}{n}\int (\delta _{n}\ast \mathcal{L}(\,\cdot
\,;N))(\lambda )m_n^{(1)}(d\lambda )-\int (\delta _{n}\ast V)(\lambda
)N(d\lambda )-Cn^{-1}\log n \\
&&\hspace{1cm} = -2\frac{n-1}{n}\mathcal{L}[N,m_n^{(1)}]-\int V(\lambda
)N(d\lambda )+O(n^{-1})-Cn^{-1}\log n \\
&&\hspace{1cm}=-\int u_{n}(\lambda ;m_n^{(1)})N(d\lambda )+O(n^{-1}\log n).
\end{eqnarray*}%
Hence, we have proved the second inequality in (\ref{n3}).

By a similar argument we derive from (\ref{n7b}), (\ref{n7}), (\ref{n4a})
and (\ref{n6}) that%
\begin{eqnarray*}
&&\hspace{-1cm}\frac{2}{\beta n }\log \int_{-1/2}^{1/2}e^{-\beta n
u_{n}(\lambda; m_n^{(1)} )/2}d\lambda \\
&&\geq -2\frac{n-1}{n}\mathcal{L}[\delta^\ast_{n}m_n^{(1)},m_n^{(1)}]-\int
(\delta _{n}\ast V)(\lambda )m_n^{(1)}(d\lambda )-Cn^{-1}\log n \\
&&\geq -2\frac{n-1}{n}\mathcal{L}[m_n^{(1)},m_n^{(1)}]-\int V(\lambda
)m_n^{(1)}(d\lambda )+O(n^{-1 })-Cn^{-1}\log n \\
&&\geq -\int u_{n}(\lambda ;m_n^{(1)})m_n^{(1)}(d\lambda )+O(n^{-1}\log n).
\end{eqnarray*}%
In view of (\ref{Phi^1}) and (\ref{F})
\begin{equation*}
F(0)-\Phi ^{(1)}[m_n^{(1)}]=-\int u_{n}(\lambda
;m_n^{(1)})m_n^{(1)}(d\lambda )-\frac{2}{\beta n}\int_{-1/2}^{1/2}e^{-\beta
n u_{n}(\lambda; m_n^{(1)} )/2}d\lambda
\end{equation*}%
and the third inequality of (\ref{n3}) follows. Lemma \ref{l:IDS.l2} is
proved.


\section{Bulk universality of local eigenvalue statistics.}

\subsection{Generalities}

Universality is an important asymptotic property of spectra of random
matrices of large size $n$. According to the property (see e.g. \cite%
{We-Co:98,Ka-Sa:99,Me:91}) the probabilistic description of eigenvalues on
the scale of typical spacing does not depend on the matrix probability law
(ensemble) in the limit $n\rightarrow \infty $ and may only depend on the
type of matrices (real symmetric, hermitian, or quaternion real in the case
of real eigenvalues and orthogonal, unitary or symplectic in the case of the
eigenvalues on the unit circle).

In a more concrete setting of the bulk of the spectrum of hermitian matrix
models (\ref{MMb}) -- (\ref{dMb}) the property can be described as follows.
Assume that the limiting Normalized Counting Measure of eigenvalues $N$ (see
e.g. Theorem \ref{t:IDS.t1} for its existence) possesses a continuous
density $\rho $ (see e.g. Theorem \ref{t:IDS.2}). Choose $\lambda _{0}$
belonging to the bulk of the support of $N$, i.e., such that $0<\rho
(\lambda _{0})< \infty$, and assume that $\rho_n$ of (\ref{Nnrho}) converges
uniformly to $\rho$ in a neighborhood of $\lambda_0$. Then we have to have
the following limiting relation for any marginal density (\ref{pnlb}) for $%
\beta =2$:
\begin{eqnarray}  \label{U.1.9}
\lim_{n\rightarrow \infty }[\rho _{n}(\lambda _{0})]^{-l}p_{l,2}^{(n)}\Big(%
\lambda _{0}+{\frac{x_{1}}{n\rho _{n}(\lambda _{0})}},...,\lambda _{0}+{%
\frac{x_{l}}{n\rho _{n}(\lambda _{0})}}\Big) =\det \big\{S(x_{1}-x_{k})\big\}%
_{j,k=1}^{l},
\end{eqnarray}%
where
\begin{equation}
S(x)={\frac{\sin \pi x}{\pi x}}.  \label{U.1.10}
\end{equation}%
In other words, the limit in the r.h.s. of (\ref{U.1.9}) should not depend
on $V$ in (\ref{MMb}) (modulo some weak conditions) for all $\lambda _{0}$
that belong to the bulk of the spectrum. Note that the r.h.s. of (\ref{U.1.9}%
) does not depend on $\lambda _{0}$.

Thus the limit (\ref{U.1.9}) for arbitrary $V$ has to coincide with that for
the archetype Gaussian Unitary Ensemble, corresponding to $%
V(\lambda)=\lambda^2/ 2$. For this case (\ref{U.1.9}) is known since the
early sixties (see \cite{Me:91} for corresponding results and discussions).

In addition, an analogous properties has to be valid for the "hole"
probability
\begin{equation}
E_{n,2}(\Delta )=\mathbf{P}\{\lambda _{l}^{(n)}\notin \Delta
,\;l=1,...,n\},\quad \Delta \subset \mathbb{R}.  \label{hole}
\end{equation}%
Namely, we have to have for any $s>0$:
\begin{equation}
\lim_{n\rightarrow \infty }E_{n,2}([\lambda _{0},\lambda _{0}+s/n\rho_n
(\lambda _{0})])=\det (1-S_{s})  \label{holim}
\end{equation}%
where $S_{s}$ is the integral operator, defined by the kernel $S(x-y)$ on
the interval $[0,s]$.

We will prove the following

\begin{theorem}
\label{t:U.t1} Consider a matrix model (\ref{MMb}) -- (\ref{dMb}) for $\beta
=2$ and assume that its potential $V$ satisfies (\ref{cond0}), $V^{\prime }$
is a Lipschitz function (see (\ref{Lip})) and there exists a closed interval $%
[a,b]\subset \sigma =\emph{supp}\,N$ such that
\begin{equation}
\sup_{\lambda \in \lbrack a,b]}|V^{\prime \prime \prime }(\lambda )|\leq
C_{1}<\infty ,\quad 0<\inf_{\lambda \in \lbrack a,b]}\rho (\lambda ).
\label{cond3}
\end{equation}%
Then for any $d>0$ the universality properties (\ref{U.1.9}) and (\ref{holim}%
) are true for any $\lambda _{0}\in \lbrack a+d,b-d)$. More precisely

\begin{itemize}
\item[(i)] (\ref{U.1.9}) is true uniformly in $(x_{1},...x_{l})$, varying on
a compact set of $\mathbb{R}^{l}$;

\item[(ii)] (\ref{holim}) is true uniformly in $s$, varying on a compact set
of $[0,\infty )$.
\end{itemize}
\end{theorem}

The theorem will be proved in this and the next subsections. An important
technical mean of the proof is a remarkable formula for all marginals (\ref%
{pnlb}) of the joint eigenvalue probability density (\ref{psymb}) for $\beta
=2$. The formula is known as the determinant formula (see e.g \cite{Me:91}
for details).

Assume that $V$ satisfies (\ref{cond0}) and consider polynomials $%
\{P_{l}^{(n)}(\lambda )\}_{l \geq 0}$ orthogonal on $\mathbb{R}$ with
respect to the weight
\begin{equation}
w_{n}(\lambda )=e^{-nV(\lambda )}.  \label{U.1.12}
\end{equation}%
We have
\begin{equation}
\int P_{l}^{(n)}(\lambda )P_{m}^{(n)}(\lambda )e^{-nV(\lambda )}d\lambda
=\delta _{l,m},  \label{U.1.13}
\end{equation}%
or, denoting
\begin{equation}
\psi _{l}^{(n)}(\lambda )=\exp \{-nV(\lambda )/2\}P_{l}^{(n)}(\lambda
),\,l=0,1,...,  \label{U.1.14}
\end{equation}%
we obtain the corresponding orthogonal functions in $L^{2}(\mathbb{R})$:%
\begin{equation}
\int \psi _{l}^{(n)}(\lambda )\psi _{m}^{(n)}(\lambda )d\lambda =\delta
_{l,m}.  \label{U.1.15}
\end{equation}%
Then marginal densities (\ref{pnlb}) have the determinant form \cite{Me:91}
\begin{equation}
p_{l,2}^{(n)}(\lambda _{1},...,\lambda _{l})={\frac{(n-l)!}{n!}}\det
\{K_{n}(\lambda _{j},\lambda _{k})\}_{j,k=1}^{l},  \label{U.1.18}
\end{equation}%
where
\begin{equation}
K_{n}(\lambda ,\mu )=\sum_{l=0}^{n-1}\psi _{l}^{(n)}(\lambda )\psi
_{l}^{(n)}(\mu ), \quad \int K_{n}(\lambda ,\nu )K_{n}(\nu, \mu) d\mu =
K_{n}(\lambda ,\mu )  \label{U.1.19}
\end{equation}%
is known as the reproducing kernel of system (\ref{U.1.14}). In particular,
\begin{equation}
\rho _{n}(\lambda ):=p_{1,2}^{(n)}(\lambda )=n^{-1}K_{n}(\lambda ,\lambda
)=n^{-1}\sum_{l=0}^{n-1}(\psi _{l}^{(n)}(\lambda ))^{2}.  \label{U.1.20}
\end{equation}%
We mention also the Christoffel-Darboux formula \cite{Sz:67}:%
\begin{equation}
K_{n}(\lambda ,\mu )=J_{n-1}^{(n)}\frac{\psi _{n}^{(n)}(\lambda )\psi
_{n-1}^{(n)}(\mu )-\psi _{n-1}^{(n)}(\lambda )\psi _{n}^{(n)}(\mu )}{\lambda
-\mu },  \label{U.3.4}
\end{equation}%
where%
\begin{equation}
J_{k}^{(n)}=\int \lambda \psi _{k}^{(n)}(\lambda )\psi _{k+1}^{(n)}(\lambda
)d\lambda ,\;k=0,1,...  \label{U.3.5}
\end{equation}%
are the off-diagonal coefficients of the Jacobi matrix, associated with
these orthogonal polynomials.

Write the hole probability as%
\begin{equation*}
E_{n,\beta }(\Delta )=\mathbf{E}\left\{ \prod\limits_{l=1}^{n}\left( 1-\chi
_{\Delta }\left( \lambda _{l}^{(n)}\right) \right) \right\} ,
\end{equation*}%
where $\chi _{\Delta }$ is the indicator of $\Delta \subset \mathbb{R}$, use
the symmetry of (\ref{psymb}) in its arguments, (\ref{U.1.18}), and the
scaling of the l.h.s. of (\ref{holim}). This yields for $\Delta=[\lambda_0,
\lambda _{0}+s/n\rho (\lambda _{0})]$:%
\begin{eqnarray}  \label{hdet}
E_{n,2}\left( \left[ \lambda _{0},\lambda _{0}+s/n\rho_n (\lambda _{0})%
\right] \right) =\det(I-K_n\chi_\Delta)\hspace{7cm} \\
=1+\sum_{l=1}^\infty\rho_n ^{-l}(\lambda _{0})\int_{[0,s]^l}dx_1\dots dx_l
\det\{K_n(\lambda _{0}+x_i/n\rho_n (\lambda _{0}), \lambda _{0}+x_j/n\rho_n
(\lambda _{0}))\}_{i,j=1}^l  \notag
\end{eqnarray}%
where $K_n$ is the integral operator with the kernel $K_n$ of (\ref{U.1.19})
and $\chi_\Delta$ is the multiplication operator by $\chi_\Delta$. In view
of (\ref{U.1.18}) and (\ref{hdet}) the proof of the universality properties (%
\ref{U.1.9}) and (\ref{holim}) for the random matrix ensemble (\ref{MMb}) --
(\ref{dMb}) with $\beta =2$ reduces in essence to the proof of the limiting
relation
\begin{equation}
\lim_{n\rightarrow \infty }(n\rho_n (\lambda _{0}))^{-1}K_{n}\left( \lambda
_{0}+x/n\rho_n (\lambda _{0}),\lambda _{0}+y/n\rho_n (\lambda _{0})\right) ={%
\frac{\sin \pi (x-y)}{\pi (x-y)}.}  \label{U.1.23}
\end{equation}%
In paper \cite{De-Co:99} the asymptotic formulas for $\psi _{n}^{(n)},\;\psi
_{n-1}^{(n)},\;J_{n-1}^{(n)}$ as $n\rightarrow \infty $ were found in the
case of a real analytic potential, and the limits (\ref{U.1.9}) and (\ref%
{holim}) were obtained by using above formulas, (\ref{U.3.4}) for in
particular. In paper \cite{Pa-Sh:97} a certain integral representation for $%
K_{n}(\lambda ,\mu )$ was used (see formula (\ref{U.2.10}) below) to obtain
the \emph{sin}-kernel of the r.h.s. of (\ref{U.1.23}) as a series in its
argument. In this paper we start from the same representation of the
reproducing kernel and derive an integro-differential equation for the limit
of the l.h.s of (\ref{U.1.23}). We then show that a unique solution of the
equation is the \emph{sin}-kernel of the r.h.s. of (\ref{U.1.23}). It turns
out that this requires weaker conditions (see Theorem \ref{t:U.t1}) than the
potential to be a real analytic function. In view of this and the importance
of the universality properties (\ref{U.1.9}) and (\ref{holim}) it seems
reasonable to present one more proof of the property.


\subsection{Proof of basic results.}

An important ingredient of our proof is the uniform convergence of $\rho_{n}$
of (\ref{Nnrho}) to $\rho $ of (\ref{sq-rep}) in a neighborhood of $%
\lambda_0 $.

\begin{theorem}
\label{t:U.t2} Under conditions of Theorem \ref{t:U.t1} we have for any$%
\;d>0 $:
\begin{equation}
\sup_{\lambda \in \lbrack a+d,b-d]}|\rho _{n}(\lambda )-\rho (\lambda )|\leq
Cn^{-2/9}  \label{U.1.25}
\end{equation}%
with some positive and finite $C$.
\end{theorem}

\textbf{Proof.} We note first again that we can assume without loss of
generality that $V$ is linear for large absolute values of its argument,
i.e., that (\ref{b_V'}) is valid (see the beginning of the proof of Theorem %
\ref{t:IDS.2}). Using in (\ref{sq-eq.1}) \ representation (\ref{U.1.18}) for
$p_{2}^{(n)}(\lambda ,\mu )$, we obtain for $z=\lambda +i\eta $, $\eta >0$:
\begin{equation}
f_{n}^{2}(z)+V^{\prime }(\lambda )f_{n}(z)+Q_{n}(\lambda ,\eta )=-\frac{1}{%
n^{2}}\int K_{n}^{2}(\lambda ,\mu )\bigg(\frac{1}{\lambda -z}-\frac{1}{\mu -z%
}\bigg)^{2}d\lambda d\mu ,  \label{p2.5}
\end{equation}%
where $f_{n}(z)$ was defined in (\ref{p2.4}), and
\begin{equation*}
Q_{n}(\lambda ,\eta )=\int \displaystyle\frac{V^{\prime }(\mu )-V^{\prime
}(\lambda )}{\mu -z}\ \rho _{n}(\mu )d\mu .
\end{equation*}%
is well defined due to (\ref{U.b_rho}), (\ref{b_V'}), and our conditions on $%
V(\lambda )$ (see Theorem \ref{t:U.t1}).

\smallskip To proceed further we need two lemmas, whose proof will be given
in the next subsection.

\begin{lemma}
\label{l:U.l1} Let $K_{n}(\lambda ,\mu )$ be defined by (\ref{U.1.19}). Then
for any $\delta >0$ we have under conditions of Theorem \ref{t:IDS.t1}:
\begin{eqnarray}
\int (\lambda -\mu )^{2}K_{n}^{2}(\lambda ,\mu )d\lambda d\mu &\leq &C,\quad
\int_{|\lambda -\mu |>\delta }K_{n}^{2}(\lambda ,\mu )d\lambda d\mu \leq
C\delta ^{-2},  \label{U.3.2} \\
\bigg|\int (\lambda -\mu )^{\alpha }K_{n}^{2}(\lambda ,\mu )d\mu \bigg| %
&\leq &C\bigg([\psi _{n-1}^{(n)}(\lambda )]^{2}+[\psi _{n}^{(n)}(\lambda
)]^{2}\bigg),\quad \alpha =1,2,  \label{U.3.3a} \\
\int_{|\lambda -\mu |>\delta }K_{n}^{2}(\lambda ,\mu )d\mu &\leq &C\delta
^{-2}\bigg([\psi _{n-1}^{(n)}(\lambda )]^{2}+[\psi _{n}^{(n)}(\lambda )]^{2}%
\bigg).  \label{U.3.3b}
\end{eqnarray}
\end{lemma}

\begin{lemma}
\label{l:U.l2} Under the conditions of Theorem~\ref{t:U.t1} we have
uniformly in $\lambda \in \lbrack a+d,b-d]$ for any$\;d>0$
\begin{equation}
\rho _{n}(\lambda )\leq C,  \label{U.b_rho}
\end{equation}%
\begin{equation}
\left\vert {\frac{d\rho _{n}(\lambda )}{d\lambda }}\right\vert \leq C\bigg(%
[\psi _{n-1}^{(n)}(\lambda )]^{2}+[\psi _{n}^{(n)}(\lambda )]^{2}\bigg)+C,
\label{U.3.21}
\end{equation}%
\begin{equation}
\int_{|\mu -\lambda |\leq n^{-1/4}}d\mu \left( \lbrack \psi _{n-1}^{(n)}(\mu
)]^{2}+[\psi _{n}^{(n)}(\mu )]^{2}\right) \leq Cn^{-1/4},  \label{U.3.24}
\end{equation}%
\begin{equation}
{\frac{1}{n}}\bigg([\psi _{n-1}^{(n)}(\lambda )]^{2}+[\psi
_{n}^{(n)}(\lambda )]^{2}\bigg)\leq Cn^{-1/8}.  \label{U.3.25}
\end{equation}
\end{lemma}

\noindent It follows then from (\ref{p2.5}) and (\ref{U.3.2}) that
\begin{equation}
f_{n}^{2}(z)+V^{\prime }(\lambda )f_{n}(z)+Q_{n}(\lambda ,\eta
)=O(n^{-2}\eta ^{-4}),  \label{p2.6}
\end{equation}%
where $\eta =|\Im z|$. Observe now that if $\lambda \in \lbrack a+d,b-d]$,
then we have for sufficiently small $\eta $ in view of (\ref{Lip}), (\ref%
{b_V'}), and (\ref{U.b_rho}):
\begin{equation*}
\begin{array}{c}
|Q_{n}(\lambda ,\eta )-Q_{n}(\lambda ,0)|\leq \eta \displaystyle\int_{|\mu
-\lambda |>d/2}\displaystyle\frac{|V^{\prime }(\mu )-V^{\prime }(\lambda
)|\rho _{n}(\mu )d\mu }{|\mu -\lambda ||(\mu -\lambda )^{2}+\eta ^{2}|^{1/2}}
\\
+C\eta \displaystyle\int_{|\mu -\lambda |<d/2}\displaystyle\frac{d\mu }{%
|(\mu -\lambda )^{2}+\eta ^{2}|^{1/2}}\leq C\eta d^{-2}+C\eta \log \eta
^{-1}.%
\end{array}%
\end{equation*}%
Besides, applying (\ref{w_conv}), we get
\begin{equation*}
Q_{n}(\lambda ,0)=Q(\lambda )+O(n^{-1/2}\log ^{1/2}n),
\end{equation*}%
where $Q$ is defined by (\ref{Q}). The last two bounds yield
\begin{equation}
Q_{n}(\lambda ,\eta )=Q(\lambda )+O(n^{-1/2}\log ^{1/2}n)+O(\eta \log \eta
^{-1}).  \label{U.2.6}
\end{equation}%
Combining (\ref{p2.6}) and (\ref{U.2.6}), we find for any $\eta \geq
n^{-3/8} $ that
\begin{equation}
\begin{array}{lll}
f_{n}(\lambda +i\eta ) & = & -\displaystyle\frac{1}{2}V^{\prime }(\lambda )+%
\bigg[V^{\prime }{}^{2}(\lambda )/4-Q(\lambda ) \\
&  & +O(\eta \log \eta ^{-1})+O(n^{-1/2}\log ^{1/2}n)+O(n^{-2}\eta ^{-4})%
\bigg]^{1/2}.%
\end{array}
\label{U.2.8}
\end{equation}%
This and (\ref{sq-rep}) yield for $\eta =n^{-4/9}$:
\begin{equation}
\pi ^{-1}\Im f_{n}(\lambda +i\eta )=\rho (\lambda )+O(n^{-2/9}).
\label{U.2.9}
\end{equation}%
On the other hand, integrating by parts and using (\ref{U.2.8}) and Lemma~%
\ref{l:U.l2}, we obtain for $\eta =n^{-4/9}$%
\begin{eqnarray*}
&&\hspace{-1.5cm}|\pi ^{-1}\Im f_{n}(\lambda +i\eta )-\rho _{n}(\lambda )| \\
&\leq &\displaystyle\frac{\eta }{\pi }\left( \int_{|\mu -\lambda |<\eta
^{1/2}}+\int_{\eta ^{1/2}<|\mu -\lambda |<d/2}\right) \frac{|\rho _{n}(\mu
)-\rho _{n}(\lambda )|}{(\mu -\lambda )^{2}+\eta ^{2}}d\mu +O(\eta ) \\
&\leq &C\int_{|\mu -\lambda |<\eta ^{1/2}}|\rho _{n}^{\prime }(\mu )|d\mu
+O(\eta ^{1/2})\leq C\eta ^{1/2}.
\end{eqnarray*}
This bound and (\ref{U.2.9}) imply (\ref{U.1.25}).

\medskip

\textbf{Proof of Theorem~\ref{t:U.t1}.} According to (\ref{U.1.18}), the
proof of validity of (\ref{U.1.9}) -- (\ref{U.1.10}) uniformly on a compact
set of $\mathbb{R}^{l}$, i.e., assertion (i) of the theorem, reduces to the
proof of validity of limiting relation (\ref{U.1.23}) for the reproducing
kernel (\ref{U.1.19}) of the orthonormal systems (\ref{U.1.14}) uniformly in
$(x,y)$ on a compact set of $\mathbb{R}^{2}$. This proof occupies the
overwhelming part of the this and the next subsections. Before presenting
the proof we will show that (\ref{U.1.23}) implies (\ref{holim}), i.e.,
assertion (ii) of the theorem. Indeed, if (\ref{U.1.23}) is valid, then we
can pass to the limit $n\rightarrow \infty $ in the integrals over $%
(x_{1},...,x_{l})$ in every term of (\ref{hdet}) and obtain the r.h.s. of (%
\ref{U.1.9}) as the integrand of every integral. We have to prove then that
the terms of (\ref{hdet}) are bounded uniformly in $n$ by terms of a
convergent series. This is based on

\begin{lemma}
\label{l:Hadpd} Let $A=\{A_{jk}\}_{j,k=1}^l$ be a positive definite $l\times
l$ matrix. Then%
\begin{equation}
\det A\leq \prod\limits_{j=1}^{l}A_{jj}.  \label{Hadpd}
\end{equation}
\end{lemma}

\noindent The lemma will be proved in the next subsection. It follows from (%
\ref{U.1.19}) that the matrix $\{K_{n}(\lambda _{j},\lambda
_{k})\}_{j,k=1}^{l}$ is positive definite. Hence we have by the above lemma,
(\ref{U.1.20}), and (\ref{U.b_rho}):%
\begin{equation*}
\det \{(n\rho _{n}(\lambda _{0}))^{-1}K_{n}(\lambda _{0}+x_{j}/n\rho
_{n}(\lambda _{0}),\lambda _{0}+x_{k}/n\rho _{n}(\lambda
_{0}))\}_{j,k=1}^{l}\leq \prod\limits_{j=1}^{l}\frac{\rho _{n}(\lambda
_{0}+x_{j}/n\rho _{n}(\lambda _{0}))}{\rho _{n}(\lambda _{0})}\leq C^{l}.
\end{equation*}%
Thus, the $l$th term of (\ref{hdet}) is bounded by $C^{l}/l!$, \ the term of
a convergent series. This allows us to pass to the limit $n\rightarrow
\infty $ in every term of (\ref{hdet}) and to obtain (\ref{holim}) in view
of (\ref{U.1.20}).

\medskip

We turn now to the proof of validity of (\ref{U.1.23}) uniformly in $(x,y)$
on a compact set of $\mathbb{R}^{2}$. This will be based on the
representation
\begin{equation}
\begin{array}{lll}
n^{-1}K_{n}(\lambda ,\mu ) & = & Q_{n,2}^{-1}e^{-n(V(\lambda )+V(\mu ))/2}
\\
& \times & \displaystyle\int \displaystyle\prod_{j=2}^{n}d\lambda
_{j}e^{-nV(\lambda _{j})}(\lambda -\lambda _{j})(\mu -\lambda _{j})%
\displaystyle\prod_{2\leq j<k\leq n}(\lambda _{j}-\lambda _{k})^{2}%
\end{array}
\label{U.2.10}
\end{equation}%
which can be derived from the well-known identities of random matrix theory
\cite{Me:91}
\begin{equation*}
\prod_{1\leq j<k\leq n}(\lambda _{j}-\lambda _{k})=\left(
\prod_{l=0}^{n-1}\gamma _{l}^{(n)}\right) ^{-1}\det \{P_{j-1}^{(n)}(\lambda
_{k})\}_{j,k=0}^{n-1},\ \ Q_{n,2}=n!\prod_{l=1}^{n}(\gamma _{l}^{(n)})^{-2},
\end{equation*}%
where $\gamma _{l}^{(n)}$ is the coefficient in front of $\lambda ^{l}$ in
the polynomial $P_{l}^{(n)}$. Using the first identity with $\lambda
_{1}=\lambda $ and $\lambda _{1}=\mu $ in the r.h.s. of (\ref{U.2.10}),
integrating the result with respect to $\lambda _{2},...,\lambda _{n}$ we
obtain the l.h.s. of (\ref{U.1.10}), in view of the orthonormality of
functions (\ref{U.1.14}).

We note again that we can assume without loss of generality that the
potential is linear for $|\lambda |>L$, where $L$ is defined in Theorem \ref%
{t:IDS.t1}, i.e., that (\ref{b_V'}) is valid. Corresponding argument is a
version of that leading to (\ref{HoldVev}) and (\ref{b_V'}). Indeed, if $%
V_{1}$ and $V_{2}$ satisfy the conditions of Proposition \ref{t:IDS.2}, then
$V(\lambda ,t)=tV_{1}(\lambda )+(1-t)V_{2}(\lambda ),\,t\in \lbrack 0,1] $
also does. Denote $\overline{N}_{n}(\cdot ,t)$ the measure (\ref{Nnrho}) and
$\rho _{n}(\cdot ,t)$ its density. Then, by using formulas (\ref{U.1.12}) --
(\ref{U.1.20}) and (\ref{U.2.10}), we obtain for the kernel $K_{n}(\lambda
,\mu ,t)$, corresponding to $V(\lambda ,t)$:
\begin{eqnarray}
\frac{\partial}{\partial t}K_{n}(\lambda ,\mu ,t) &=&-\frac{n}{2}(\delta
V(\lambda )+\delta V(\mu ))K_{n}(\lambda ,\mu ,t)  \label{ddtK} \\
&&+\int \delta V(\nu )K_{n}(\lambda ,\nu ,t)K_{n}(\mu ,\nu ,t)d\nu ,  \notag
\end{eqnarray}%
where $\delta V=V_{1}-V_{2}$. Now, if $V_{1}(\lambda )=V_{2}(\lambda )$, $%
|\lambda |\leq L$, then in view of the inequality (see (\ref{U.1.19}) and (%
\ref{U.1.20}))
\begin{equation}
K_{n}^{2}(\lambda ,\mu )\leq K_{n}(\lambda ,\lambda )K_{n}(\mu ,\mu
)=n^{2}\rho _{n}(\lambda )\rho _{n}(\mu )  \label{KSch}
\end{equation}%
we obtain for $\lambda ,\mu \in \lbrack a+d,b-d]\subset (a,b)$ ($d>0$):
\begin{equation*}
\left\vert \frac{\partial}{\partial t}K_{n}(\lambda ,\mu ,t)\right\vert \leq
n^{3}\rho _{n}^{1/2}(\lambda,t )\rho _{n}^{1/2}(\mu,t )\int_{|\lambda
|>L}|\delta V(\nu )|\rho _{n}(\nu,t )d\nu ,
\end{equation*}%
It follows from (\ref{U.b_rho}), (\ref{exp_est}), and (\ref{cond0}) that the
r.h.s. of this inequality is $O(e^{-nc})$ uniformly in $t\in [ 0,1]$ (cf (%
\ref{dNbo})). Hence, the limit (\ref{U.1.23}) for a given potential,
satisfying the condition of the theorem, is the same as that for the
potential, coinciding with the given for $|\lambda |\leq L$ and linear for,
say $|\lambda |\geq L+1 $.

Now  take some $\lambda _{0}\in \lbrack a+d,b-d]$, where $[a,b]$ is
defined in (\ref{cond3}), and denote
\begin{equation}
\mathcal{K}_{n}(x,y)=n^{-1}K_{n}(\lambda _{0}+x/n,\lambda _{0}+y/n).
\label{cal-K}
\end{equation}%
We have from (\ref{U.1.19}) -- (\ref{U.1.20}), (\ref{U.b_rho}), and (\ref%
{KSch}):
\begin{equation}
\int \mathcal{K}_{n}(x,z)\mathcal{K}_{n}(z,y)dz=\mathcal{K}_{n}(x,y),\quad
\mathcal{K}_{n}^{2}(x,y)\leq \mathcal{K}_{n}(x,x)\mathcal{K}_{n}(y,y),
\label{cKSch}
\end{equation}%
\begin{equation}
\mathcal{K}_{n}(x,x)=\rho _{n}(\lambda _{0}+x/n)\leq C<\infty ,\quad |%
\mathcal{K}_{n}(x,y)|\leq C<\infty ,\;x,y=o(n).  \label{ocKr}
\end{equation}%
Then, differentiating (\ref{U.2.10}) with respect to $x$, we get (cf (\ref%
{ddtK}))
\begin{equation}
\begin{array}{lll}
\displaystyle\frac{\partial }{\partial x}\mathcal{K}_{n}(x,y) & = & -%
\displaystyle\frac{1}{2}V^{\prime }(\lambda _{0}+x/n)\mathcal{K}_{n}(x,y) \\
&  & +\displaystyle\int \displaystyle\frac{\mathcal{K}_{n}(x^{\prime
},x^{\prime })\mathcal{K}_{n}(x,y)-\mathcal{K}_{n}(x,x^{\prime })\mathcal{K}%
_{n}(x^{\prime },y)}{x-x^{\prime }}dx^{\prime }.%
\end{array}
\label{tU.1}
\end{equation}%
We have the following lemma that will be proved in the next subsection.

\begin{lemma}
\label{l:d1} Denote
\begin{equation*}
D(\lambda ) = \frac{V^{\prime }(\lambda )}{2}-\frac{1}{n}\int \frac{%
K_{n}(\mu ,\mu )d\mu }{\lambda -\mu }.
\end{equation*}%
Then under conditions of Theorem \ref{t:U.t1} we have uniformly in any $%
[a+d,b-d]\subset (a,b)$:
\begin{equation}
|D(\lambda )|\leq Cn^{-1/4}\log n.  \label{ld.1}
\end{equation}
\end{lemma}

\noindent The lemma yields
\begin{equation*}
\begin{array}{l}
\displaystyle\frac{1}{2}V^{\prime }(\lambda _{0}+x/n)\mathcal{K}_{n}(x,y)-%
\displaystyle\int \displaystyle\frac{\mathcal{K}_{n}(x^{\prime },x^{\prime })%
\mathcal{K}_{n}(x,y)}{x-x^{\prime }}dx^{\prime } \\
\hskip3cm=D(\lambda _{0}+x/n)\mathcal{K}_{n}(x,y)=O(n^{-1/4}\log n).%
\end{array}%
\end{equation*}%
This allows us to rewrite (\ref{tU.1}) as
\begin{equation}
\frac{\partial }{\partial x}\mathcal{K}_{n}(x,y)=-\int \frac{\mathcal{K}%
_{n}(x,x^{\prime })\mathcal{K}_{n}(x^{\prime },y)}{x-x^{\prime }}dx^{\prime
}+O(n^{-1/4}\log n).  \label{tU.2}
\end{equation}%
Denote%
\begin{equation}
\mathcal{L}=\log n.  \label{cal-L}
\end{equation}%
For $|x|,|y|\leq \mathcal{L}$ we can restrict integration in (\ref{tU.2}) by
the domain $|x^{\prime }|\leq 2\mathcal{L}$, replacing $O(n^{-1/4}\log n)$
by $O(\mathcal{L}^{-1})$, where $\mathcal{L}$ is defined by (\ref{cal-L}).
This follows from the bound
\begin{equation}
\bigg|\int_{|x^{\prime }|>2\mathcal{L}}\frac{\mathcal{K}_{n}(x,x^{\prime })%
\mathcal{K}_{n}(x^{\prime },y)}{x-x^{\prime }}dx^{\prime }\bigg|\leq
\mathcal{L}^{-1}\bigg(\int \mathcal{K}_{n}^{2}(x,x^{\prime })dx^{\prime
}\int \mathcal{K}_{n}^{2}(y,x^{\prime })dx^{\prime }\bigg)^{1/2}\leq C%
\mathcal{L}^{-1}.  \label{b_out_L}
\end{equation}%
and (\ref{cKSch}) -- (\ref{ocKr}).

We will use now the following assertion that will be proved in the next
subsection.

\begin{lemma}
\label{l:U.l4} Under conditions of Theorem \ref{t:U.t1} we have uniformly in
$|x|,|y|<\mathcal{L},\;$ $\lambda _{0}\in \lbrack a+d,b-d]$
\begin{equation}
\left\vert \frac{\partial }{\partial x}\mathcal{K}_{n}(x,y)+\displaystyle%
\frac{\partial }{\partial y}\mathcal{K}_{n}(x,y)\right\vert \leq C\left(
n^{-1/8}+|x-y|n^{-2}\right) ,  \label{U.3.34a}
\end{equation}%
\begin{equation}
|\mathcal{K}_{n}(x,y)-\mathcal{K}_{n}(0,y-x)|\leq C|x|\left(
n^{-1/8}+|x-y|n^{-2}\right) ,  \label{U.3.34b}
\end{equation}%
\begin{equation}
\bigg|\displaystyle\frac{\partial }{\partial x}\mathcal{K}_{n}(x,y)\bigg|%
\leq C,\quad \displaystyle\int_{|x|\leq \mathcal{L}}dx\bigg|\displaystyle%
\frac{\partial }{\partial x}\mathcal{K}_{n}(x,y)\bigg|^{2}\leq C.
\label{b_der}
\end{equation}
\end{lemma}

\noindent Denote
\begin{eqnarray}
\mathcal{K}_{n}^{\ast }(x) &=&\mathcal{K}_{n}(x,0)\mathbf{1}_{|x|\leq
\mathcal{L}}+\mathcal{K}_{n}(\mathcal{L},0)(1+\mathcal{L}-x)\mathbf{1}_{%
\mathcal{L}<x\leq \mathcal{L}+1}  \label{k^*} \\
&+&\mathcal{K}_{n}(-\mathcal{L},0)(1+\mathcal{L}+x)\mathbf{1}_{-\mathcal{L}%
-1\leq x<-\mathcal{L}},  \notag
\end{eqnarray}%
and observe that if we set $x=0$ in (\ref{tU.2}) and take $|y|\leq \mathcal{L%
}/3$, then similarly to (\ref{b_out_L}) we can restrict integration to $%
|x^{\prime }|\leq 2\mathcal{L}/3$ in the obtained relation, adding $O(%
\mathcal{L}^{-1})$. This and Lemma~ \ref{l:U.l4} lead to the equation
\begin{equation}
\displaystyle\frac{\partial }{\partial y}\mathcal{K}_{n}^{\ast
}(y)=\int_{|x^{\prime }|\leq 2\mathcal{L}/3}\frac{\mathcal{K}_{n}^{\ast
}(x^{\prime })\mathcal{K}_{n}^{\ast }(y-x^{\prime })}{x^{\prime }}dx^{\prime
}+r_{n}(y)+O(\mathcal{L}^{-1}),  \label{tU.4}
\end{equation}%
where
\begin{equation*}
r_{n}(y)=\int_{|x^{\prime }|\leq 2\mathcal{L}/3}\frac{\mathcal{K}%
_{n}(0,x^{\prime })(\mathcal{K}_{n}(x^{\prime },y)-\mathcal{K}%
_{n}(0,y-x^{\prime }))}{x^{\prime }}dx^{\prime },
\end{equation*}%
and assuming that $|y|\leq \mathcal{L}/3$ we have by Lemma ~\ref{l:U.l4}
\begin{equation*}
r_{n}(y)=O(n^{-1/8}\log n).
\end{equation*}%
Now, using the bound similar to (\ref{b_out_L}), we can replace in (\ref%
{tU.4}) the integral over $|x^{\prime }|\leq 2\mathcal{L}/3$ by the integral
over the whole real line. Besides, on the basis of Lemma~\ref{l:U.l4} and (%
\ref{cKSch}) -- (\ref{ocKr}), we obtain
\begin{equation}
\displaystyle\int |\mathcal{K}_{n}^{\ast }(x)|^{2}dx\leq \displaystyle\int |%
\mathcal{K}_{n}(x,0)|^{2}dx+C^{\prime }\leq C,\quad \displaystyle\int \bigg|%
\displaystyle\frac{d}{dx}\mathcal{K}_{n}^{\ast }(x)\bigg|^{2}dx\leq C.
\label{l2nor}
\end{equation}%
Consider the Fourier transform
\begin{equation*}
\widehat{\mathcal{K}}_{n}^{\ast }(p)=\int \mathcal{K}_{n}^{\ast
}(x)e^{ipx}dx,
\end{equation*}%
where the integral is defined in the $L^{2}(\mathbb{R})$ sense, and write $%
\mathcal{K}_{n}^{\ast }(x)$ as
\begin{equation}
\mathcal{K}_{n}^{\ast }(x)=(2\pi )^{-1}\int \widehat{\mathcal{K}}_{n}^{\ast
}(p)e^{-ipy}dp.  \label{tU.5}
\end{equation}%
Then we have from (\ref{U.1.20}) and (\ref{U.1.25}):
\begin{equation}
\int \widehat{\mathcal{K}}_{n}^{\ast }(p)dp=2\pi \rho (\lambda _{0})+o(1),
\label{Kprho}
\end{equation}%
and from (\ref{l2nor}) and the Parseval equation:%
\begin{equation}
\int p^{2}|\widehat{\mathcal{K}}_{n}^{\ast }(p)|^{2}dp\leq C.  \label{tU.5a}
\end{equation}%
It follows from (\ref{U.1.19}) and (\ref{cal-K}) that the kernel $\mathcal{K}%
_{n}(x,y)$ is positive definite:
\begin{equation*}
\int_{-\mathcal{L}}^{\mathcal{L}}\mathcal{K}_{n}(x,y)f(x)\overline{f}%
(y)dxdy\geq 0,\quad f\in L_{2}(\mathbb{R}),
\end{equation*}%
and by (\ref{U.3.34b}) we have for any $f\in L_{2}(\mathbb{R})$:
\begin{equation}
\int \widehat{\mathcal{K}}_{n}^{\ast }(p)|\hat{f}(p)|^{2}dp\geq
-C||f||_{L^{2}(\mathbb{R})}^{2}(n^{-1/8}\log ^{4}n+O(\mathcal{L}^{-1})).
\label{tU.6}
\end{equation}%
Furthermore, the Parseval equation and (\ref{U.3.34b}) yield
\begin{equation}
\int |\widehat{\mathcal{K}}_{n}^{\ast }(p)-\widehat{\mathcal{K}}_{n}^{\ast
}(-p)|^{2}dp=2\pi \int |\mathcal{K}_{n}^{\ast }(x)-\mathcal{K}_{n}^{\ast
}(-x)|^{2}dx\leq Cn^{-1/8}\log ^{3}n.  \label{tU.6a}
\end{equation}%
We write now by definition of the singular integral
\begin{equation}
\int \frac{\mathcal{K}_{n}^{\ast }(x^{\prime })\mathcal{K}_{n}^{\ast
}(y-x^{\prime })}{x^{\prime }}dx^{\prime }=\lim_{\varepsilon \rightarrow
+0}\int dx^{\prime }\mathcal{K}_{n}^{\ast }(x^{\prime })\mathcal{K}%
_{n}^{\ast }(y-x^{\prime })\Re (x^{\prime }+i\varepsilon )^{-1}.
\label{tU.6b}
\end{equation}%
In view of the formula
\begin{equation*}
\int e^{ipx}\Re (x^{\prime }+i\varepsilon )^{-1}dx=\pi ie^{-\varepsilon |p|}%
\hbox{sgn}\,p
\end{equation*}%
and the Parseval equation we can write the r.h.s. of (\ref{tU.6b}) as
\begin{multline}
\frac{1}{2\pi }\lim_{\varepsilon \rightarrow +0}\int dpdp^{\prime }\widehat{%
\mathcal{K}}_{n}^{\ast }(p)\widehat{\mathcal{K}}_{n}^{\ast }(p^{\prime
})e^{-ipy}\hbox{sign}(p-p^{\prime })e^{-\varepsilon |p-p^{\prime }|}
\label{tU.6c} \\
=\frac{i}{2\pi }\int dpe^{-ipy}\widehat{\mathcal{K}}_{n}^{\ast
}(p)\int_{0}^{p}\widehat{\mathcal{K}}_{n}^{\ast }(p^{\prime })dp^{\prime }-%
\frac{i}{2\pi }\int dpe^{-ipy}\widehat{\mathcal{K}}_{n}^{\ast
}(p)\int_{0}^{\infty }(\widehat{\mathcal{K}}_{n}^{\ast }(p^{\prime })-%
\widehat{\mathcal{K}}_{n}^{\ast }(-p^{\prime }))dp^{\prime }.
\end{multline}%
Note that the both integrals are absolutely convergent because $\widehat{%
\mathcal{K}}_{n}^{\ast }\in L^{1}(\mathbb{R})$ by (\ref{tU.5a}). Since the
Schwarz inequality and (\ref{tU.5a}) imply the bound
\begin{eqnarray*}
\left\vert \int_{0}^{\infty }(\widehat{\mathcal{K}}_{n}^{\ast }(p^{\prime })-%
\widehat{\mathcal{K}}_{n}^{\ast }(-p^{\prime }))dp^{\prime }\right\vert
&\leq &\left\vert \int_{0}^{\mathcal{L}^{2}}(\hat{\mathcal{K}}_{n}^{\ast
}(p^{\prime })-\widehat{\mathcal{K}}_{n}^{\ast }(-p^{\prime }))dp^{\prime
}\right\vert +\int_{|p|>\mathcal{L}^{2}}|\widehat{\mathcal{K}}_{n}^{\ast
}(p^{\prime })|dp^{\prime } \\
&\leq &\mathcal{L}\left( \int |\widehat{\mathcal{K}}_{n}^{\ast }(p^{\prime
})-\widehat{\mathcal{K}}_{n}^{\ast }(-p^{\prime })|^{2}dp^{\prime }\right)
^{1/2}+C\mathcal{L}^{-1},
\end{eqnarray*}%
we get from (\ref{tU.6a}) - (\ref{tU.6c}) uniformly in $|y|<\mathcal{L}/3$
\begin{equation*}
\int \frac{\mathcal{K}_{n}^{\ast }(x^{\prime })\mathcal{K}_{n}^{\ast
}(y-x^{\prime })}{x^{\prime }}dx^{\prime }=\frac{i}{2\pi }\int dp\widehat{%
\mathcal{K}}_{n}^{\ast }(p)e^{-ipy}\int_{0}^{p}\widehat{\mathcal{K}}%
_{n}^{\ast }(p^{\prime })dp^{\prime }+O(\mathcal{L}^{-1}).
\end{equation*}%
This allows us to transform (\ref{tU.4}) into the following asymptotic
relation, valid for $|y|\leq \mathcal{L}/3$:
\begin{equation}
\int \widehat{\mathcal{K}}_{n}^{\ast }(p)\bigg(\int_{0}^{p}\widehat{\mathcal{%
K}}_{n}^{\ast }(p^{\prime })dp^{\prime }-p\bigg)e^{-ipy}dp=O(\mathcal{L}%
^{-1}).  \label{tU.7}
\end{equation}%
Now consider the functions
\begin{equation}
F_{n}(p)=\int_{0}^{p}\widehat{\mathcal{K}}_{n}^{\ast }(p^{\prime
})dp^{\prime }.  \label{FnKc}
\end{equation}%
Since $p\widehat{\mathcal{K}}_{n}^{\ast }(p)\in L^{2}(\mathbb{R})$, the
sequence $\{F_{n}(p)\}$ consists of functions that are of uniformly bounded
variation,  uniformly bounded and equicontinuous on $\mathbb{R}$. Thus $%
\{F_{n}(p)\}$ is a compact family with respect to the uniform convergence.
Hence, the limit $F$ of any subsequence $\{F_{n_{k}}\}$ possesses the
properties:

\begin{itemize}
\item[(a)] $F$ is bounded and continuous;

\item[(b)] $F(p)=-F(-p)$ (see (\ref{tU.6}));

\item[(c)] $F(p)\leq F(p^{\prime })$, if $p\leq p^{\prime }$ (see (\ref{tU.6}%
));

\item[(d)] $F(+\infty )-F(-\infty )=2\pi \rho (\lambda _{0})$ (see (\ref%
{Kprho}));

\item[(e)] $F$ satisfies the following equation, valid for any smooth
function $g$ of compact support (see (\ref{tU.7})):
\begin{equation}
\int (F(p)-p)g(p)dF(p)=0.  \label{tU.8}
\end{equation}
\end{itemize}
The last property implies that $F(p)=p$ or $F(p)=\hbox{const}$,
hence it follows from (a) -- (c) that
\begin{equation*}
F(p)=p\,\mathbf{1}_{|p|\leq p_{0}}+\rho ^{\ast }\,\hbox{sign}(p)\,\mathbf{1}%
_{|p|\geq p_{0}},
\end{equation*}%
where $p_{0}=\pi \rho (\lambda _{0})$ by (d).

We conclude that (\ref{tU.8}) is uniquely soluble, thus the sequence $%
\{F_{n}\}$ converges uniformly on any compact to the above $F$. This and (%
\ref{FnKc}) imply the weak convergence of the sequence $\{\mathcal{K}%
_{n}^{\ast }\}$ to the function $\ $
\begin{equation*}
\mathcal{K}^{\ast }(x)=\frac{\sin (\pi \rho (\lambda _{0})x)}{\pi \rho
(\lambda _{0})x}.
\end{equation*}%
But weak convergence combined with (\ref{ocKr}) and (\ref{b_der}) implies
the uniform convergence of $\{\mathcal{K}_{n}^{\ast }\}$ to $\mathcal{K}%
^{\ast }$ on any interval. Now, using Lemma~\ref{l:U.l4}, we obtain that we
have uniformly in $(x,y)$, varying on a compact set of $\mathbb{R}^2$
\begin{equation*}
\lim_{n\rightarrow \infty }\mathcal{K}_{n}(x,y)=\mathcal{K}^{\ast }(x-y).
\end{equation*}%
Recalling (\ref{U.1.9}), (\ref{U.1.18}), (\ref{U.1.23}), (\ref{U.1.25}), and
(\ref{cal-K}) we conclude that Theorem~\ref{t:U.t1} is proved.

\subsection{Auxiliary results for Theorem \protect\ref{t:U.t1}}

\textbf{Proof of Lemma \ref{l:U.l1}.} By using (\ref{U.1.15}) and the
Christoffel-Darboux formula (\ref{U.3.4}) -- (\ref{U.3.5}) we get for the
r.h.s. of (\ref{U.3.2})
\begin{equation}
\int (\lambda -\mu )^{2}K_{n}^{2}(\lambda ,\mu )d\lambda d\mu
=2(J_{n-1}^{(n)})^{2}.  \label{U.3.10}
\end{equation}%
Besides, (\ref{exp_est}) and (\ref{U.1.19}) -- (\ref{U.1.20}) imply the
bound
\begin{equation}
\lbrack \psi _{l}^{(n)}(\lambda )]^{2}\leq n\rho _{n}(\lambda )\leq n\exp
\{-CnV(\lambda )\},\quad |\lambda |\geq L,\quad l=0,1,\dots n-1,
\label{U.3.11}
\end{equation}%
and then we have by (\ref{U.3.5}) that%
\begin{equation}
|J_{n-1}^{(n)}|\leq C.  \label{U.3.12}
\end{equation}%
This bound, (\ref{cond0}) and (\ref{U.3.10}) imply (\ref{U.3.2}). Similar
argument and equation (\ref{U.3.4}) yield
\begin{equation}  \label{U.3.10a}
\int (\lambda -\mu )K_{n}^{2}(\lambda ,\mu )d\mu =J_{n-1}^{(n)}\psi
_{n-1}^{(n)}(\lambda )\psi _{n}^{(n)}(\lambda ).
\end{equation}%
Now (\ref{U.3.3a}) for $\alpha =1$ follows from this identity and (\ref%
{U.3.12}). The case $\alpha =2$ in the l.h.s. of (\ref{U.3.3a}) can be
proved similarly and (\ref{U.3.3b}) follows from (\ref{U.3.3a}) with $\alpha
=2$. Lemma \ref{l:U.l1} is proved.

\medskip

\textbf{Proof of Lemma~\ref{l:U.l2}.} We start from the simple identity
\begin{equation*}
\displaystyle\frac{d\rho _{n}(\lambda )}{d\lambda }=\displaystyle\frac{d\rho
_{n}(\lambda +t)}{dt}\bigg|_{t=0}.
\end{equation*}%
Changing variables in the integral (\ref{pnlb}) to $\lambda _{i}-t=\mu _{i}$%
, $i=2,...n$, we rewrite $\rho _{n}(\lambda +t)$ as
\begin{equation*}
\rho _{n}(\lambda +t)=Q_{n,2}^{-1}\int e^{-nV(\lambda +t)}\displaystyle%
\prod_{i>j\geq 2}^{n}(\mu _{i}-\mu _{j})^{2}\displaystyle%
\prod_{j=2}^{n}e^{-nV(t+\mu _{j})}(\lambda -\mu _{j})^{2}d\mu _{j}.
\end{equation*}%
Hence, after differentiating with respect to $t$ and setting $t=0$ in the
result we get
\begin{equation}
\begin{array}{rcl}
\displaystyle\frac{d\rho _{n}(\lambda )}{d\lambda } & = & -nV^{\prime
}(\lambda )\rho _{n}(\lambda )-n(n-1)\displaystyle\int V^{\prime }(\mu
)p_{n,2}^{(2)}(\lambda ,\mu )d\mu  \\
& = & -V^{\prime }(\lambda )K_{n}(\lambda ,\lambda )-\displaystyle\int
V^{\prime }(\mu )(K_{n}(\lambda ,\lambda )K_{n}(\mu ,\mu )-K_{n}^{2}(\lambda
,\mu ))d\mu ,%
\end{array}
\label{U.3.22}
\end{equation}%
where $p_{n,2}^{(2)}(\lambda ,\lambda _{2})$ is defined by (\ref{pnlb}) and
we used also (\ref{U.1.18}) for $l=2$. Integrating this relation and using (%
\ref{U.1.19}) we obtain
\begin{equation*}
\int V^{\prime }(\mu )K_{n}(\mu ,\mu )d\mu =0.
\end{equation*}%
This, (\ref{U.3.22}), and (\ref{U.1.19}) yield
\begin{equation}
\rho _{n}^{\prime }(\lambda )=\int (V^{\prime }(\mu )-V^{\prime }(\lambda
))K_{n}^{2}(\lambda ,\mu )d\mu .  \label{U.3.23}
\end{equation}%
We split this integral in two parts corresponding to the intervals $|\mu
-\lambda |>d/2$ and $|\mu -\lambda |\leq d/2$, and use (\ref{Lip}), (\ref{b_V'}), and (\ref%
{U.3.3b}) with $\delta =d/2$ for the former integral. In the latter
integral we write
\begin{equation*}
V^{\prime }(\mu )-V^{\prime }(\lambda )=(\mu -\lambda )V^{\prime \prime
}(\lambda )+{\frac{(\mu -\lambda )^{2}}{2}}V^{\prime \prime \prime }(\xi )
\end{equation*}%
for some $\xi $ depending on $\lambda $ and $\mu $ and use Lemma \ref{l:U.l1}
and condition (\ref{cond3}) of Theorem \ref{t:U.t1}. Combining the bounds
for these two integrals, we obtain (\ref{U.3.21}). To obtain (\ref{U.b_rho})
we use (\ref{triv_id}) for $v=\rho _{n}$ and (\ref{U.3.21}) in the first
integral of (\ref{triv_id}) and (\ref{inrho}) in the second.

To prove (\ref{U.3.24}) and (\ref{U.3.25}) we introduce the probability
density
\begin{equation}
p_{n}^{-}(\lambda _{1},...,\lambda _{n-1})={\frac{1}{Q_{n,2}^{-}}}%
\prod_{j=1}^{n-1}e^{-nV(\lambda _{j})}\prod_{1\leq j<k\leq n-1}(\lambda
_{j}-\lambda _{k})^{2}.  \label{U.3.26}
\end{equation}%
The difference of this density from density (\ref{psymb}) written for $n-1$
variables $\lambda _{1},...,\lambda _{n-1}$ is that in the former we have
the factor $n$ in the exponent while in the latter we would have $n-1$. We
have analogously to (\ref{U.1.18}) for $l=1$ and (\ref{U.1.20}):
\begin{equation}
\rho _{n}^{-}(\lambda ):={\frac{n-1}{n}}\int p_{n}^{-}(\lambda ,\lambda
_{2},...,\lambda _{n-1})d\lambda _{2}...d\lambda _{n-1}={\frac{1}{n}}%
\sum_{j=0}^{n-2}(\psi _{j}^{(n)}(\lambda ))^{2},  \label{U.3.27}
\end{equation}%
\noindent thus
\begin{equation}
(\psi _{n-1}^{(n)}(\lambda ))^{2}=n(\rho _{n}(\lambda )-\rho
_{n}^{-}(\lambda )).  \label{U.3.28}
\end{equation}%
Furthermore, by using an analog of identity (\ref{p2.6}) for the probability
density $p_{n}^{-}$, we obtain the asymptotic relation
\begin{equation}
(f_{n}^{-}(z))^{2}+\int {\frac{V^{\prime }(\mu )\rho _{n}^{-}(\mu )}{\mu -z}}%
d\mu =O(n^{-2}\eta ^{-4})  \label{U.3.29}
\end{equation}%
for the Stieltjes transform $f_{n}^{-}$ of $\rho _{n}^{-}$ and $z=\lambda
+i\eta ,\ \eta >0$. Denote
\begin{equation}
\Delta _{n}(z):=n(f_{n}(z)-f_{n}^{-}(z))=\int \frac{(\psi _{n-1}^{(n)}(\mu
))^{2}}{\mu -z}d\mu ,  \label{U.3.30}
\end{equation}%
subtract (\ref{U.3.29}) from (\ref{p2.6}) and multiply the result by $n$.
This yields:
\begin{equation*}
\Delta _{n}(z)(f_{n}(z)+f_{n}^{-}(z))+\int \frac{V^{\prime }(\mu )}{\mu -z}%
(\psi _{n-1}^{(n)}(\mu ))^{2}d\mu =O(n^{-2}\eta ^{-4}).
\end{equation*}%
For $z=\lambda +in^{-1/4}$ this relation takes the form
\begin{equation*}
\Delta _{n}(z)(f_{n}(z)+f_{n}^{-}(z)-V^{\prime }(\lambda ))=\int \frac{%
V^{\prime }(\lambda )-V^{\prime }(\mu )}{\mu -z}(\psi _{n-1}^{(n)}(\mu
))^{2}d\mu +O(1).
\end{equation*}%
Since $\Im f_{n}^{-}(z)\Im z>0,\;\Im z>0$, we can write in view of (\ref%
{cond3}) and (\ref{U.2.8})
\begin{equation*}
0<\Im \Delta _{n}(\lambda +in^{-1/4})\leq \left( \frac{1}{\Im f_{n}(z)}\int
\frac{V^{\prime }(\lambda )-V^{\prime }(\mu )}{\lambda -\mu }(\psi
_{n-1}^{(n)}(\mu ))^{2}d\mu +O(1)\right) \leq C,
\end{equation*}%
and then (\ref{U.3.30}) for $z=\lambda +in^{-1/4}$ yields:%
\begin{eqnarray}
\int_{|\mu -\lambda |\leq n^{-1/4}}(\psi _{n-1}^{(n)}(\mu ))^{2}d\mu  &\leq
&2n^{-1/2}\int \frac{(\psi _{n-1}^{(n)}(\mu ))^{2}}{(\mu -\lambda
)^{2}+n^{-1/2}}d\mu   \label{U.3.31} \\
&=&2n^{-1/4}\Im \Delta _{n}(\lambda +in^{-1/4})\leq Cn^{-1/4}.  \notag
\end{eqnarray}%
Hence, we have proved (\ref{U.3.24}).

To prove (\ref{U.3.25}) for $\psi _{n-1}^{(n)}$ we need two elementary
facts. The first is the inequality for a differentiable function $%
u:[a_{1},b_{1}]\rightarrow \mathbb{C}$:
\begin{equation}
||u||_{\infty }^{2}\leq 2||u||_{2}\,||u^{\prime
}||_{2}+(b_{1}-a_{1})^{-1}||u||_{2}^{2},  \label{Ag}
\end{equation}%
where $||\dots ||_{\infty }$ and $||\dots ||_{2}$ are the uniform and the $%
L^{2}$ - norm in $[a_{1},b_{1}]$. The inequality (a simple case of the
Sobolev inequalities) follows easily from (\ref{triv_id}) with $v=u^{2}$ and
the Schwarz inequality.

The second fact is the identity
\begin{equation*}
\int \left( \frac{d}{d\mu }\psi _{n-1}^{(n)}(\mu )\right) ^{2}d\mu =\int {%
\frac{n^{2}}{4}}V^{\prime 2}(\mu )\Big(\psi _{n-1}^{(n)}(\mu)\Big)^{2}d\mu,
\end{equation*}%
that follows from (\ref{U.1.13}) -- (\ref{U.1.14}) and the integration by
part, taking into account that $P_{k}^{(n)}$ is orthogonal to its second
derivative, a polynomials of degree $k-2$. The identity, (\ref{b_V'}), and (%
\ref{U.1.15}) yield the bound
\begin{equation*}
\int \left( \frac{d}{d\mu }\psi _{n-1}^{(n)}(\mu )\right) ^{2}d\mu \leq
Cn^{2}.
\end{equation*}%
This, (\ref{Ag}) for $u=\psi _{n-1}^{(n)}$, $[a_{1},b_{1}]=[\lambda
-n^{1/4},\lambda +n^{1/4}]$, and (\ref{U.3.31}) yield (\ref{U.3.25}) for $%
\psi _{n-1}^{(n)}$.

To prove an analogous bounds for $\psi _{n}^{(n)}$ we repeat the above
argument for the probability density (cf (\ref{U.3.26}))
\begin{equation*}
p_{n}^{+}(\lambda _{1},...,\lambda _{n+1})={\frac{1}{Q_{n,2}^{+}}}%
\prod_{1\leq jn+1}e^{-nV(\lambda _{j})}\prod_{1\leq j<k\leq n+1}(\lambda
_{j}-\lambda _{k})^{2},
\end{equation*}%
setting
\begin{equation*}
\rho _{n}^{+}(\lambda ):=\frac{n+1}{n}\int p_{n}^{+}(\lambda ,\lambda
_{2},...,\lambda _{n+1})d\lambda _{2}...d\lambda _{n+1}=\frac{1}{n}%
\sum_{j=0}^{n}[\psi _{j}^{(n)}(\mu )]^{2},
\end{equation*}%
so that $[\psi _{n}^{(n)}(\lambda )]^{2}=n(\rho _{n}^{+}(\lambda )-\rho
_{n}(\lambda ))$ (cf (\ref{U.3.28})). Lemma \ref{l:U.l2} is proved.

\medskip \textbf{Proof of Lemma \ref{l:Hadpd}.} \ Since $A$ is positive
definite there exists a positive definite $B$ such that $A=B^{2}$. We have
then by the Hadamard inequality:%
\begin{equation*}
\det A=\det B^{2}\leq \prod\limits_{j=1}^{l}\sum_{k=1}^{l}|B_{jk}|^{2}.
\end{equation*}%
By definition of $B$ the sum in the r.h.s. is $(B^{2})_{jj}=A_{jj}$ and we
obtain the assertion of lemma.

\medskip \textbf{Proof of Lemma \ref{l:d1}.} According to (\ref{U.2.8}) we
have for $f_{n}$, defined by (\ref{p2.4})
\begin{equation}
|\Re f_{n}(\lambda +i\eta )+V^{\prime }(\lambda )/2|\leq Cn^{-3/8}\log
n,\quad \eta =n^{-3/8}.  \label{ld1.2}
\end{equation}%
On the other hand, using (\ref{p2.4}), integrating by parts the difference $%
\Re f_{n}(\lambda +i\eta )-\Re f_{n}(\lambda +i0)$, written via the r.h.s.
of (\ref{p2.4}), and using (\ref{U.3.21}) and (\ref{U.b_rho}), we obtain%
\begin{eqnarray}
&&\hspace{-1cm}\left\vert \Re f_{n}(\lambda +i\eta )-\int \frac{\rho
_{n}(\mu )d\mu }{\mu -\lambda }\right\vert =\frac{1}{2}\left\vert \int_{|\mu
-\lambda |\leq d/2}\log (1+\eta ^{2}|\mu -\lambda |^{-2})\rho _{n}^{\prime
}(\mu )d\mu \right\vert +O(\eta )  \notag  \label{ld1.3} \\
&\leq &C\int_{|\mu -\lambda |\leq d/2}\log (1+\eta ^{2}|\mu -\lambda
|^{-2})\left( (\psi _{n-1}^{(n)}(\mu ))^{2}+(\psi _{n}^{(n)}(\mu
))^{2}\right) d\mu +O(\eta )  \notag \\
&=&C(I_{1}+I_{2}+I_{3})+O(\eta ),  \notag
\end{eqnarray}%
where $d$ is given in the formulation of Theorem \ref{t:U.t1} and $I_{1}$, $%
I_{2}$, and $I_{3}$ correspond to the integrals over $|\lambda -\mu |\leq
n^{-2}$, $n^{-2}\leq |\lambda -\mu |\leq n^{-1/4}$, and $n^{-1/4}\leq
|\lambda -\mu |\leq d/2$ respectively. Using (\ref{U.3.25}) for $I_{1}$ and (%
\ref{U.3.24}) for $I_{2}$, we get
\begin{equation*}
I_{1}\leq Cn^{-2}\log n\leq Cn^{-1},\quad I_{2}\leq Cn^{-1/4}\log n.
\end{equation*}%
Besides, for $\eta =n^{-3/8}$ and $|\lambda -\mu |>n^{-1/4}$ we have the
inequality $\log (1+\eta ^{2}/|\mu -\lambda |^{2})=O(n^{-1/4})$, thus $%
I_{3}=O(n^{-1/4})$. Hence, we obtain from (\ref{ld1.3}) that
\begin{equation*}
\bigg|\Re f_{n}(\lambda +i\eta )-\int \frac{\rho _{n}(\mu )d\mu }{\mu
-\lambda }\bigg|\leq Cn^{-1/4}\log n.
\end{equation*}%
This inequality and (\ref{ld1.2}) prove Lemma~\ref{l:d1}.

\medskip \textbf{Proof of Lemma~\ref{l:U.l4}.} \ To simplify notations we
denote
\begin{equation}
\lambda _{x}=\lambda _{0}+(x-tx)/n,\ \ \ \lambda _{y}=\lambda _{0}+(y-tx)/n.
\label{lxly}
\end{equation}%
Then, repeating almost literally the derivation of (\ref{U.3.23}), we get
the formula
\begin{equation}\label{dtKxy}
\frac{d}{dt}K_{n}(\lambda _{x},\lambda _{y})=x\int K_{n}(\lambda
_{x},\lambda )K_{n}(\lambda _{y},\lambda )\left( {\frac{1}{2}}V^{\prime
}(\lambda _{x})+{\frac{1}{2}}V^{\prime }(\lambda _{y})-V^{\prime }(\lambda
)\right) d\lambda .
\end{equation}%
To estimate the r.h.s. of the formula we split the integral in two parts
corresponding to the intervals $|\lambda -\lambda _{0}|>d/2$ and $|\lambda
-\lambda _{0}|\leq d/2$, where $d=\max \{\lambda _{0}-a,b-\lambda _{0}\}$,
and for the former integral we use the inequality $2K_{n}(\lambda ,\lambda
_{x})K_{n}(\lambda ,\lambda _{y})\leq K_{n}^{2}(\lambda ,\lambda
_{x})+K_{n}^{2}(\lambda ,\lambda _{y})$ and then (\ref{U.3.3b}) with $\delta
=d/2$ and (\ref{b_V'}). In the latter integral we write
\begin{eqnarray*}
&&\hspace{-1.5cm}V^{\prime }(\lambda )-{\frac{1}{2}}V^{\prime }(\lambda
_{x})-{\frac{1}{2}}V^{\prime }(\lambda _{y}) \\
&=&{\frac{1}{2}}(\lambda -\lambda _{x})V^{\prime \prime }(\lambda _{x})+{%
\frac{1}{2}}(\lambda -\lambda _{y})V^{\prime \prime }(\lambda _{y})+O\left(
(\lambda -\lambda _{x})^{2}+(\lambda -\lambda _{y})^{2}\right)  \\
&=&{\frac{1}{2}}(\lambda -\lambda _{x})V^{\prime \prime }(\lambda _{x})+{%
\frac{1}{2}}(\lambda -\lambda _{y})V^{\prime \prime }(\lambda _{y})+O\left(
(\lambda -\lambda _{x})(\lambda -\lambda _{y})+\displaystyle\frac{|x-y|^{2}}{%
n^{2}}\right) .
\end{eqnarray*}%
The Christoffel-Darboux formula (\ref{U.3.4}) yields (cf (\ref{U.3.10a}))
\begin{equation*}
\int K_{n}(\lambda _{x},\lambda )K_{n}(\lambda _{y},\lambda )(\lambda
-\lambda _{x})d\lambda =-J_{n-1}^{(n)}\psi _{n}^{(n)}(\lambda _{x})\psi
_{n-1}^{(n)}(\lambda _{y}).
\end{equation*}%
Hence%
\begin{eqnarray*}
&&\hspace{-1.5cm}\int_{|\lambda -\lambda _{0}|\leq d}K_{n}(\lambda
_{x},\lambda )K_{n}(\lambda _{y},\lambda )(\lambda -\lambda _{x})d\lambda  \\
&=&\bigg(\int -\int_{|\lambda -\lambda _{0}|\geq d}\bigg)K_{n}(\lambda
_{x},\lambda )K_{n}(\lambda _{y},\lambda )(\lambda -\lambda _{x,y})d\lambda
\\
&=&-J_{n-1}^{(n)}\psi _{n}^{(n)}(\lambda _{x})\psi _{n-1}^{(n)}(\lambda
_{y})-I_{d},
\end{eqnarray*}%
where $I_{d}$ can be estimated by using again (\ref{U.3.3b}) and an argument
similar to that in (\ref{dtKxy}). Similar formulas are valid for $(\lambda
-\lambda _{y})$ in the integrals. Besides, we have by Schwarz inequality,%
\begin{eqnarray*}
&&\hspace{-1.5cm}\left\vert \int K_{n}(\lambda _{x},\lambda )K_{n}(\lambda
_{y},\lambda )(\lambda -\lambda _{x})(\lambda -\lambda _{y})d\lambda
\right\vert  \\
&\leq &\left[ \int K_{n}^{2}(\lambda _{x},\lambda )(\lambda -\lambda
_{x})^{2}d\lambda \int K_{n}^{2}(\lambda _{y},\lambda )(\lambda -\lambda
_{y})^{2}d\lambda \right] ^{1/2}.
\end{eqnarray*}%
Using (\ref{U.3.3a}) for the r.h.s. of the last inequality and the above
estimates for the integrals with $(\lambda -\lambda _{x})$ and $(\lambda
-\lambda _{y})$ we obtain from (\ref{dtKxy})
\begin{eqnarray}
&&\hspace{-1.5cm}\left\vert \frac{d}{dt}K_{n}(\lambda _{x},\lambda
_{y})\right\vert   \label{dtKp} \\
&\leq &C|x|\left( (\psi _{n}^{(n)}(\lambda _{x}))^{2}+(\psi
_{n-1}^{(n)}(\lambda _{x}))^{2}+(\psi _{n}^{(n)}(\lambda _{y}))^{2}+(\psi
_{n-1}^{(n)}(\lambda _{y}))^{2}+\frac{|x-y|^{2}}{n}\right) .  \notag
\end{eqnarray}%
The bound, the finite increment formula, and (\ref{U.3.25}) imply (\ref%
{U.3.34b}). On the other hand, we have
\begin{equation*}
\frac{\partial }{\partial x}\mathcal{K}_{n}(x,y)+\displaystyle\frac{\partial
}{\partial y}\mathcal{K}_{n}(x,y)=-x^{-1}n^{-1}\displaystyle\frac{d}{dt}%
K_{n}(\lambda _{x},\lambda _{y})\bigg|_{t=0}.
\end{equation*}%
Combining this with (\ref{dtKp}) and (\ref{U.3.25}), we obtain (\ref{U.3.34a}%
).

Note, that (\ref{U.3.34b}) with $\lambda _{0}+x_{1}/n$ instead of $\lambda
_{0}$ and $y=x=x_{2}-x_{1}$ leads to the bound, valid for any $|x_{1,2}|<nd/2
$:
\begin{equation}
|\mathcal{K}_{n}(x_{1},x_{1})-\mathcal{K}_{n}(x_{2},x_{2})|\leq
Cn^{-1/8}|x_{1}-x_{2}|.  \label{Kxx}
\end{equation}%
To prove (\ref{b_der}) we first show that for any $|x|\leq nd/2$ we have the
bound
\begin{equation}
\displaystyle\int_{-1}^{1}\displaystyle\frac{\mathcal{K}_{n}(x,x)\mathcal{K}%
_{n}(x+t,x+t)-\mathcal{K}_{n}^{2}(x+t,x)}{t^{2}}dt\leq C.  \label{U.3.35}
\end{equation}%
To this end consider the quantity%
\begin{equation*}
W=\left\langle \prod_{i=2}^{n}\left\vert 1-{\frac{1}{n^{2}(\lambda
_{i}-\lambda _{0})^{2}}}\right\vert \right\rangle ,
\end{equation*}%
where the symbol $<...>$ denotes the operation $\mathbf{E}\{\delta (\lambda
_{1}-\lambda _{0})\dots \}$ and $\mathbf{E}\{...\}$ is the expectation with
respect to the measure (\ref{MMb}) -- (\ref{dMb}) for $\beta =2$. By Schwarz
inequality $W^{2}$ is bounded from above by the product of integrals
\begin{equation*}
Z_{n}^{-1}\int e^{-nV(\lambda _{0})}\prod_{2\leq j<k\leq n}(\lambda
_{j}-\lambda _{k})^{2}\prod_{2\leq j\leq n}(\lambda _{0}+\sigma -\lambda
_{j})^{2}e^{-nV(\lambda _{j})}d\lambda _{j}
\end{equation*}%
for $\sigma =\pm 1/n$. Besides, $n(V(\lambda _{0})-V(\lambda _{0}+\sigma ))$
is bounded in $n$ because of condition (\ref{cond3}). Replacing $V(\lambda
_{0})$ by $V(\lambda _{0}+\sigma )$ in the above integral and using (\ref%
{p2.1}) and (\ref{U.b_rho}) we can write the bound
\begin{equation}
W\leq C\cdot \rho _{n}^{1/2}(\lambda _{0}+{1/n})\rho _{n}^{1/2}(\lambda _{0}-%
{1/n})\leq C_{1}.  \label{U.3.36}
\end{equation}%
On the other hand, $W$ can be represented as
\begin{equation*}
W=\left\langle \displaystyle\prod_{i=2}^{n}(\phi _{1}(\lambda _{i})+\phi
_{2}(\lambda _{i}))\right\rangle =\displaystyle\sum_{k=0}^{n-1}\left(
_{\,\,\,k}^{n-1}\right) \left\langle \displaystyle\prod_{i=2}^{k+1}\phi
_{1}(\lambda _{i})\prod_{i=k+2}^{n}\phi _{2}(\lambda _{i})\right\rangle ,
\end{equation*}%
where
\begin{equation*}
\phi _{1}(\lambda )=\frac{(1-n^{2}(\lambda -\lambda _{0})^{2})^{2}}{%
n^{2}(\lambda -\lambda _{0})^{2}}\mathbf{1}_{n|\lambda -\lambda _{0}|<1},
\end{equation*}%
and
\begin{equation*}
\phi _{2}(\lambda )=\left( 1-n^{2}(\lambda -\lambda _{0})^{2}\right) \mathbf{%
1}_{n|\lambda -\lambda _{0}|<1}+\left( 1-n^{-2}|\lambda -\lambda
_{0}|^{-2}\right) \mathbf{1}_{n|\lambda -\lambda _{0}|>1}.
\end{equation*}%
Since $0\leq \phi _{2}(\lambda )\leq 1$ and $\phi _{1}(\lambda )\geq 0$ we
get from the term $k=1$ of the above representation:
\begin{equation}
W\geq (n-1)\int d\lambda \phi _{1}(\lambda )\left\langle \delta (\lambda
_{2}-\lambda )\exp \left\{ \sum_{i=3}^{n}\log \phi _{2}(\lambda
_{i})\right\} \right\rangle .  \label{U.3.37}
\end{equation}%
Now the Jensen inequality implies%
\begin{eqnarray}
&&\hspace{-1.5cm}\left\langle \delta (\lambda _{2}-\lambda )\exp \left\{
\sum_{i=3}^{n}\log \phi _{2}(\lambda _{i})\right\} \right\rangle
\label{U.3.38} \\
&\geq &{\exp \left\{ \left\langle \delta (\lambda _{2}-\lambda
)\sum_{i=3}^{n}\log \phi _{2}(\lambda _{i})[p_{2,2}^{(n)}(\lambda
_{0},\lambda )]^{-1}\right\rangle \right\} }  \notag \\
&=&\exp \left\{ (n-2)\int \log \phi _{2}(\lambda ^{\prime
})p_{3,2}^{(n)}(\lambda _{0},\lambda ,\lambda ^{\prime })d\lambda ^{\prime
}[p_{2,2}^{(n)}(\lambda _{0},\lambda )]^{-1}\right\} ,  \notag
\end{eqnarray}%
where $\langle \delta (\lambda _{2}-\lambda )\rangle =p_{2,2}^{(n)}(\lambda
_{0},\lambda )$ and $p_{3,2}^{(n)}(\lambda _{0},\lambda ,\lambda ^{\prime })$
are the second and the third marginal densities, specified by (\ref{pnlb})
for $\beta =2$. According to (\ref{U.1.18}) for $l=2,3$ we have
\begin{eqnarray}
&&\hspace{-1cm}p_{3,2}^{(n)}(\lambda _{0},\lambda ,\lambda ^{\prime })=%
\displaystyle\frac{n}{n-2}\rho _{n}(\lambda ^{\prime
})\,p_{2,2}^{(n)}(\lambda _{0},\lambda )  \label{U.3.39} \\
&&\hspace{-0.7cm}+\frac{2K_{n}(\lambda _{0},\lambda )K_{n}(\lambda
_{0},\lambda ^{\prime })K_{n}(\lambda ,\lambda ^{\prime })-K_{n}(\lambda
_{0},\lambda _{0})K_{n}^{2}(\lambda ,\lambda ^{\prime })-K_{n}(\lambda
,\lambda )K_{n}^{2}(\lambda _{0},\lambda ^{\prime })}{n(n-1)(n-2)}.  \notag
\end{eqnarray}%
In view of (\ref{KSch}) we can write
\begin{equation*}
\begin{array}{lll}
2K_{n}(\lambda _{0},\lambda )K_{n}(\lambda _{0},\lambda ^{\prime
})K_{n}(\lambda ^{\prime },\lambda ) & \leq  & 2K_{n}^{1/2}(\lambda
_{0},\lambda _{0})K_{n}^{1/2}(\lambda ,\lambda )|K_{n}(\lambda _{0},\lambda
^{\prime })||K_{n}(\lambda ^{\prime },\lambda )| \\
& \leq  & K_{n}(\lambda _{0},\lambda _{0})K_{n}^{2}(\lambda ^{\prime
},\lambda )+K_{n}(\lambda ,\lambda )K_{n}^{2}(\lambda _{0},\lambda ^{\prime
}).%
\end{array}%
\end{equation*}%
Thus the second term in the r.h.s. of (\ref{U.3.39}) is non-positive and we
obtain the bound
\begin{equation*}
p_{3,2}^{(n)}(\lambda _{0},\lambda ,\lambda ^{\prime })\leq \displaystyle%
\frac{n}{n-2}\rho _{n}(\lambda ^{\prime })\,p_{2,2}^{(n)}(\lambda
_{0},\lambda ).
\end{equation*}%
Hence, taking into account that $\log \phi _{2}(\lambda )\leq 0$ and $\rho
_{n}(\lambda )\leq C,\;\lambda \in \lbrack a+d,b-d]$ (see (\ref{U.b_rho})),
restricting the integration in (\ref{U.3.37}) by the interval $|\lambda
-\lambda _{0}|\leq n^{-1}$, using (\ref{U.3.38})--(\ref{U.3.39}), and
recalling the definitions of $\phi _{1,2}$, we have
\begin{equation}
\begin{array}{lll}
W & \geq  & (n-1)\displaystyle\int d\lambda \phi _{1}(\lambda
)p_{2,2}^{(n)}(\lambda _{0},\lambda )\exp \left\{ n\displaystyle\int \rho
_{n}(\lambda ^{\prime })\log \phi _{2}(\lambda ^{\prime })d\lambda ^{\prime
}\right\}  \\
& \geq  & \displaystyle\frac{n-1}{n}\displaystyle\int_{-1}^{1}\displaystyle%
\frac{(1-t^{2})^{2}}{t^{2}}p_{2,2}^{(n)}(\lambda _{0},\lambda _{0}+t/n)dt \\
& \times  & \exp \left\{ -C\left( \displaystyle\int_{0}^{1}|\log
(1-y^{2})|dy+\displaystyle\int_{1}^{\infty }\log (1-y^{-2})dy\right)
-C\right\} .%
\end{array}
\label{U.3.40}
\end{equation}%
It is easy now to derive (\ref{U.3.35}) for $x=0$ from (\ref{U.3.40}) and (%
\ref{U.3.36}). Then, replacing $\lambda _{0}$ by $\lambda _{0}+x/n$, we
obtain the same inequality for any $|x|\leq nd/2$.

Now we are ready to prove (\ref{b_der}). According to (\ref{tU.2}), we have%
\begin{eqnarray}
\bigg|\frac{\partial }{\partial x}\mathcal{K}_{n}(x,y)\bigg| &=&\bigg|\bigg(%
\int_{|x-x^{\prime }|<1}+\int_{|x-x^{\prime }|\geq 1}\bigg)\frac{\mathcal{K}%
_{n}(x,x^{\prime })\mathcal{K}_{n}(x^{\prime },y)}{x-x^{\prime }}dx^{\prime }%
\bigg|+o(1)  \label{b_der1} \\
&\leq &|I_{1}(x,y)|+|I_{2}(x,y)|+o(1).  \notag
\end{eqnarray}%
By (\ref{cKSch}) and (\ref{ocKr}) we have
\begin{equation*}
|I_{2}(x,y)|\leq \mathcal{K}_{n}^{1/2}(y,y)\mathcal{K}_{n}^{1/2}(x,x)\leq C.
\end{equation*}%
To estimate $I_{1}$ denote
\begin{equation}
t_{1}^{\ast }=\inf \{t>0:\mathcal{K}_{n}(x\pm t,x)\leq \rho _{n}(\lambda
_{0})/2\},\quad t^{\ast }=\min \{t_{1}^{\ast },1\}.  \label{t^*}
\end{equation}%
Then we can write%
\begin{eqnarray*}
I_{1}(x,y) &=&\left( \int_{|x-x^{\prime }|<t^{\ast }}+\int_{t^{\ast }\leq
|x-x^{\prime }|<1}\right) \frac{\mathcal{K}_{n}(x,x^{\prime })\mathcal{K}%
_{n}(x^{\prime },y)-\mathcal{K}_{n}(x,x)\mathcal{K}_{n}(x,y)}{x-x^{\prime }}%
dx^{\prime } \\
&=&I_{1}^{\prime }+I_{1}^{\prime \prime }.
\end{eqnarray*}%
In view of (\ref{cKSch}) -- (\ref{ocKr}) we have
\begin{equation*}
|I_{1}^{\prime \prime }|\leq C|\log t^{\ast }|.
\end{equation*}%
On the other hand, using the Schwarz inequality and (\ref{cal-K}), we obtain
the bound
\begin{equation}
\begin{array}{l}
|\mathcal{K}_{n}(x,z)-\mathcal{K}_{n}(x^{\prime },z)|^{2}=\left\vert n^{-1}%
\displaystyle{\sum_{k=0}^{n}}(\psi _{k}^{(n)}(\lambda _{0}+\frac{x}{n})-\psi
_{k}^{(n)}(\lambda _{0}+\frac{x^{\prime }}{n}))\psi _{k}^{(n)}(\lambda _{0}+%
\frac{x^{\prime }}{n}))\right\vert ^{2} \\
\hspace{3.8cm}\leq \left( \mathcal{K}_{n}(x,x)+\mathcal{K}_{n}(x^{\prime
},x^{\prime })-2\mathcal{K}_{n}(x^{\prime },x)\right) \mathcal{K}_{n}(z,z)
\\
\hspace{1cm}=\left( (\mathcal{K}_{n}^{1/2}(x,x)-\mathcal{K}%
_{n}^{1/2}(x^{\prime },x^{\prime }))^{2}+(\mathcal{K}_{n}^{1/2}(x,x)\mathcal{%
K}_{n}^{1/2}(x^{\prime },x^{\prime })-\mathcal{K}_{n}(x^{\prime },x)\right)
\mathcal{K}_{n}(z,z).%
\end{array}
\label{b1}
\end{equation}%
In view of (\ref{Kxx}) and (\ref{ocKr}) the contribution of the first term
in the parentheses of the r.h.s. of (\ref{b1}) is bounded by $%
Cn^{-1/4}|x-x^{\prime }|^{2}$. Furthermore, write the expression in the
parentheses of the second term as
\begin{equation*}
\mathcal{K}_{n}^{1/2}(x,x)\mathcal{K}_{n}^{1/2}(x^{\prime },x^{\prime })-%
\mathcal{K}_{n}(x^{\prime },x)=\frac{\mathcal{K}_{n}(x,x)\mathcal{K}%
_{n}(x^{\prime },x^{\prime })-\mathcal{K}_{n}(x^{\prime },x)}{\mathcal{K}%
_{n}^{1/2}(x,x)\mathcal{K}_{n}^{1/2}(x^{\prime },x^{\prime })+\mathcal{K}%
_{n}(x^{\prime },x)}
\end{equation*}%
and use the inequality $\mathcal{K}_{n}(x^{\prime },x)>\frac{1}{2}\rho
_{n}(\lambda _{0})>C$, valid for $|x-x^{\prime }|\leq t^{\ast }$. We obtain
the bound
\begin{eqnarray}
|\mathcal{K}_{n}(x,z)-\mathcal{K}_{n}(x^{\prime },z)|^{2} &\leq &C\left(
\mathcal{K}_{n}(x,x)+\mathcal{K}_{n}(x^{\prime },x^{\prime })-2\mathcal{K}%
_{n}(x^{\prime },x)\right)   \label{b1.a} \\
&\leq &C_{1}\bigg(\frac{|x-x^{\prime }|^{2}}{n^{1/4}}+\mathcal{K}_{n}(x,x)%
\mathcal{K}_{n}(x^{\prime },x^{\prime })-\mathcal{K}_{n}^{2}(x^{\prime },x)%
\bigg).  \notag
\end{eqnarray}%
Thus we have from (\ref{b1.a}) with $z=x,y$, (\ref{ocKr}), and the Schwarz
inequality
\begin{eqnarray}
|I_{1}^{\prime }| &=&\bigg|\displaystyle\int_{|x-x^{\prime }|\leq t^{\ast }}%
\displaystyle\frac{\mathcal{K}_{n}(x^{\prime },x)\mathcal{K}_{n}(x^{\prime
},y)-\mathcal{K}_{n}(x,x)\mathcal{K}_{n}(x,y)}{x-x^{\prime }}dx^{\prime }%
\bigg|  \label{b2} \\
&\leq &C\displaystyle\int_{|x-x^{\prime }|\leq t^{\ast }}\displaystyle\frac{|%
\mathcal{K}_{n}(x^{\prime },y)-\mathcal{K}_{n}(x,y)|}{|x-x^{\prime }|}%
dx^{\prime }+C\displaystyle\int_{|x-x^{\prime }|\leq t^{\ast }}\displaystyle%
\frac{|\mathcal{K}_{n}(x,x^{\prime })-\mathcal{K}_{n}(x,x)|}{|x-x^{\prime }|}%
dx^{\prime }  \notag \\
&\leq &C_{1}t^{\ast }+C_{1}(t^{\ast })^{1/2}\bigg(\displaystyle%
\int_{|x-x^{\prime }|\leq t^{\ast }}\displaystyle\frac{|\mathcal{K}_{n}(x,x)%
\mathcal{K}_{n}(x^{\prime },x^{\prime })-\mathcal{K}_{n}^{2}(x^{\prime },x)|%
}{|x-x^{\prime }|^{2}}dx^{\prime }\bigg)^{1/2}\leq C_{2}(t^{\ast })^{1/2},
\notag
\end{eqnarray}%
where we used (\ref{U.3.35}) to estimate the last integral. Now, on the
basis of (\ref{b_der1}) -- (\ref{b2}) and the finite increment formula, we
have that
\begin{equation*}
C_{1}<\rho _{n}(\lambda _{0})/2\leq |\mathcal{K}_{n}(x+t^{\ast },x)-\mathcal{%
K}_{n}(x,x)|\leq C_{2}\left( (t^{\ast })^{3/2}+t^{\ast }|\log t^{\ast
}|\right) .
\end{equation*}%
We conclude that the inequality $|t^{\ast }|\geq d^{\ast }$ is valid with
some $n$-independent $d^{\ast }$, hence, repeating derivations of (\ref%
{b_der1}) -- (\ref{b2}) with $d^{\ast }$ instead of $t^{\ast }$, we obtain
the first inequality of (\ref{b_der}).

To prove the second inequality in (\ref{b_der}) we observe first that we
have by (\ref{U.3.34a}):
\begin{eqnarray*}
\int_{|x|\leq \mathcal{L}}\bigg|\frac{\partial }{\partial x}\mathcal{K}%
_{n}(x,y)\bigg|^{2}dx=\int_{|x|\leq \mathcal{L}}\bigg|\frac{\partial }{%
\partial y}\mathcal{K}_{n}(x,y)\bigg|^{2}dx+o(1),\;|y|\leq \mathcal{L}.
\end{eqnarray*}%
Then we rewrite an analog of (\ref{tU.2}) for $\frac{\partial }{\partial y}%
\mathcal{K}_{n}(x,y)$ as
\begin{eqnarray*}
\frac{\partial }{\partial y}\mathcal{K}_{n}(x,y)&=&\left( \int_{|x^{\prime
}-y|\leq d^{\ast }}+\int_{|x^{\prime }|\leq 2\mathcal{L}}\mathbf{1}%
_{|x^{\prime }-y|\geq d^{\ast }}\right) \frac{\mathcal{K}_{n}(x,x^{\prime })%
\mathcal{K}_{n}(x^{\prime },y)}{y-x^{\prime }}dx^{\prime }+O(\mathcal{L}%
^{-1}) \\
&=&I_{1}(x,y)+I_{2}(x,y)+O(\mathcal{L}^{-1}).
\end{eqnarray*}
Since in $I_{1}$ the interval of integration is symmetric with respect to $y$
we can write%
\begin{eqnarray*}
I_{1}(x,y) &=&\int_{|x^{\prime }-y|\leq d^{\ast }}\frac{(\mathcal{K}%
_{n}(x,x^{\prime })-\mathcal{K}_{n}(x,y))\mathcal{K}_{n}(x^{\prime },y)}{%
y-x^{\prime }}dx^{\prime } \\
&+&\int_{|x^{\prime }-y|\leq d^{\ast }}\frac{\mathcal{K}_{n}(x,y)(\mathcal{K}%
_{n}(x^{\prime },y)-\mathcal{K}_{n}(y,y))}{y-x^{\prime }}dx^{\prime }.
\end{eqnarray*}
Then we have by the Schwarz inequality and (\ref{ocKr})%
\begin{eqnarray*}
I_{1}^{2}(x,y) &\leq &2d^{\ast }C\int_{|x^{\prime }-y|\leq d^{\ast }}\frac{(%
\mathcal{K}_{n}(x,x^{\prime })-\mathcal{K}_{n}(x,y))^{2}dx^{\prime }}{%
(y-x^{\prime })^{2}} \\
&+&2d^{\ast }\mathcal{K}_{n}^{2}(x,y)\int_{|x^{\prime }-y|\leq d^{\ast }}%
\frac{(\mathcal{K}_{n}(x^{\prime },y)-\mathcal{K}_{n}(y,y))^{2}}{%
(y-x^{\prime })^{2}}dx^{\prime }.
\end{eqnarray*}
Now (\ref{cKSch}) and (\ref{ocKr}) lead to the bound
\begin{equation*}
\begin{array}{lll}
\displaystyle\int I_{1}^{2}(x,y)dx & \leq & 2d^{\ast }C\displaystyle%
\int_{|x^{\prime }-y|\leq d^{\ast }}dx^{\prime }\frac{\mathcal{K}%
_{n}(x^{\prime },x^{\prime })+\mathcal{K}_{n}(y,y)-2\mathcal{K}%
_{n}(x^{\prime },y)}{(y-x^{\prime })^{2}} \\
& + & 2d^{\ast }C\displaystyle\int_{|x^{\prime }-y|\leq d^{\ast }}dx^{\prime
}\frac{(\mathcal{K}_{n}(x^{\prime },y)-\mathcal{K}_{n}(y,y))^{2}}{%
(y-x^{\prime })^{2}}.%
\end{array}%
\end{equation*}%
Using the second inequality of (\ref{b1.a}) for the numerator in the first
integral and the first inequality of (\ref{b1.a}) for the numerator in the
second integral and then (\ref{U.3.35}), we obtain that the integral of $%
I_{1}^{2}(x,y)$ with respect to $x$ is bounded for $|y|\leq \mathcal{L}$.

To prove the same $I_{2}$ we use (\ref{cKSch}) -- (\ref{ocKr}) to write%
\begin{eqnarray*}
\int I_{2}^{2}(x,y)dx &\leq &\int_{|x^{\prime }|,|x^{\prime \prime }|\leq 2%
\mathcal{L}}\mathbf{1}_{|x^{\prime }-y|>d^{\ast }}\mathbf{1}_{|x^{\prime
\prime }-y|>d^{\ast }}\frac{\mathcal{K}_{n}(y,x^{\prime })\mathcal{K}%
_{n}(x^{\prime },x^{\prime \prime })\mathcal{K}_{n}(x^{\prime \prime },y)}{%
(y-x^{\prime })(y-x^{\prime \prime })}dx^{\prime }dx^{\prime \prime } \\
&\leq &C\int_{|x^{\prime }|,|x^{\prime \prime }|\leq 2\mathcal{L}}\mathbf{1}%
_{|x^{\prime }-y|>d^{\ast }}\mathbf{1}_{|x^{\prime \prime }-y|>d^{\ast
}}\left( \frac{\mathcal{K}_{n}^{2}(y,x^{\prime })}{(y-x^{\prime \prime })^{2}%
}+\frac{\mathcal{K}_{n}^{2}(y,x^{\prime \prime })}{(y-x^{\prime })^{2}}%
\right) dx^{\prime }dx^{\prime \prime } \\
&\leq &2C(d^{\ast })^{-1}\mathcal{K}_{n}(y,y)=O(1).
\end{eqnarray*}%
The above bounds for integrals of $I_{1}^{2}$ and $I_{2}^{2}$ prove the
second inequality in (\ref{b_der}). Lemma \ref{l:U.l4} is proved.

\bigskip \noindent \textbf{Acknowledgements} The final version of the paper
was written during the authors stay at the H. Poincar\'e Institute (Paris)
in the frameworks of the trimester "Phenomena in High Dimensions". We are
grateful to the Organizers of the trimester for hospitality and the CNRS and
the Marie Curie Network "Phenomena in High Dimensions" for financial support.

\end{document}